\documentclass[journal]{IEEEtran}
\makeatletter
\def\endthebibliography{%
	\def\@noitemerr{\@latex@warning{Empty `thebibliography' environment}}%
	\endlist
}
\makeatother

\newcommand{\RomanNumeralCaps}[1]
{\MakeUppercase{\romannumeral #1}}

\usepackage{cite}

\usepackage[pdftex]{graphicx} 

\usepackage[caption=false,font=footnotesize]{subfig}

\usepackage{amsmath}
\usepackage{amsfonts}
\usepackage{bm}
\usepackage{filecontents}
\usepackage{amssymb}
\usepackage{color}

\usepackage{epsfig}

\usepackage{verbatim}

\usepackage{url}

\usepackage{algorithm, tabularx}

\usepackage{lettrine}

\usepackage{lipsum}

\usepackage{siunitx}

\usepackage{soul,color}

\usepackage{mathrsfs}

\usepackage{array}

\usepackage[inline]{enumitem}

\usepackage{dblfloatfix}

\usepackage{float}

\usepackage{algorithm}

\usepackage[noend]{algpseudocode}

\usepackage{optidef}

\usepackage{fontenc}

\usepackage{hhline}

\usepackage{setspace}
\usepackage{multirow}
\usepackage{array}
\newcolumntype{L}[1]{>{\raggedright\let\newline\\\arraybackslash\hspace{0pt}}m{#1}}
\newcolumntype{C}[1]{>{\centering\let\newline\\\arraybackslash\hspace{0pt}}m{#1}}
\newcolumntype{R}[1]{>{\raggedleft\let\newline\\\arraybackslash\hspace{0pt}}m{#1}}

\newcommand{\norm}[1]{\left\lVert#1\right\rVert}

\begin{document}
	
	\title{A Generalized Pointing Error Model for\\ FSO Links with Fixed-Wing UAVs for 6G: \\Analysis and Trajectory Optimization}
	%\title{A Generalized Model of FSO Pointing Error for UAVs with Asymmetric Jitter:\\ Analysis and System Optimization}
	%\title{A New Model of FSO Pointing Error for UAVs with Asymmetric Jitter: Analysis and System Optimization}

	%Outage Probability of One-Shot Coarse Pointing with \\RF Lens Antenna Array for Free Space Optics

	\author{Hyung-Joo Moon,~\IEEEmembership{Graduate Student Member,~IEEE}, Chan-Byoung Chae,~\IEEEmembership{Fellow,~IEEE}, \\Kai-Kit Wong,~\IEEEmembership{Fellow,~IEEE}, and Mohamed-Slim Alouini,~\IEEEmembership{Fellow,~IEEE}
		\thanks{This work was supported in part by Institute of Information \& communications Technology Planning \& Evaluation (IITP) grant funded by the Korea government (MSIT) (2019-0-00685) and in part by the National Research Foundation of Korea (NRF) grant funded by the Korea government (MSIT) (2022R1A5A1027646). (\textit{Corresponding author: C.-B. Chae.})} % ACK: FSO, ERC
		\thanks{H.-J. Moon and C.-B. Chae are with the School of Integrated Technology, Yonsei University, Seoul 03722, South Korea (e-mail: \{moonhj, cbchae\}@yonsei.ac.kr).}
		\thanks{K.-K. Wong is with the Department of Electronic and Electrical Engineering, University College London, London WC1E 7JE, UK (e-mail: kai-kit.wong@ucl.ac.uk). He is also affiliated with Yonsei Frontier Lab., Yonsei University, Seoul 03722, South Korea.}
		\thanks{M.-S. Alouini is with the Computer, Electrical, and Mathematical Science and Engineering Division, King Abdullah University of Science and Technology, Thuwal 23955, Saudi Arabia (e-mail: slim.alouini@kaust.edu.sa).}% <-this % stops a space
	}
	
	%\markboth{IEEE Wireless Communications Letters,~Vol.~XX, No.~XX, XXX~2020}
	{}
	%{Shell \MakeLowercase{\textit{et al.}}: Bare Demo of IEEEtran.cls for Journals}
	
	%This work was supported in part by the Institute for Information \& communications Technology Promotion (IITP) grant funded by the Korea government (MSIP) under Grant 1711094030, and in part by the. (\textit{Corresponding author: Chan-Byoung Chae.

	\maketitle
	
	\begin{abstract}
	
%%%%%%%%%%%%%%%%%%%%%%%%%%%%%%%%%%%%%%%%%%%%%%Generated

%A demand for massive connectivity and increasing data traffic of Beyond 5G (B5G)/6G networks has prompted researchers to utilize a vertical layer of space to create a three-dimensional network called a non-terrestrial network (NTN).
Free-space optical (FSO) communication is a promising solution to support wireless backhaul links in emerging 6G non-terrestrial networks. At the link level, pointing errors in FSO links can significantly impact capacity, making accurate modeling of these errors essential for both assessing and enhancing communication performance. In this paper, we introduce a novel model for FSO pointing errors in unmanned aerial vehicles (UAVs) that incorporates three-dimensional (3D) jitter, including roll, pitch, and yaw angle jittering. We derive a probability density function for the pointing error angle based on the relative position and posture of the UAV to the ground station. This model is then integrated into a trajectory optimization problem designed to maximize energy efficiency while meeting constraints on speed, acceleration, and elevation angle. Our proposed optimization method significantly improves energy efficiency by adjusting the UAV’s flight trajectory to minimize exposure to directions highly affected by jitter. The simulation results emphasize the importance of using UAV-specific 3D jitter models in achieving accurate performance measurements and effective system optimization in FSO communication networks. Utilizing our generalized model, the optimized trajectories achieve up to 11.8$\%$ higher energy efficiency compared to those derived from conventional Gaussian pointing error models.

	\end{abstract}


	\IEEEpeerreviewmaketitle

	\section{Introduction}
	
	%\subsection{Background and Motivation}
	
\IEEEPARstart{C}{ommunication} resources are still predominantly concentrated on terrestrial networks, even with the development of 5G communication networks. To meet the demands of massive connectivity and increasing data traffic of B5G/6G, researchers are utilizing the vertical space to create a three-dimensional (3D) network known as a non-terrestrial network (NTN). This network includes satellites, high-altitude platforms (HAPs), low-altitude platforms (LAPs), and ground nodes~\cite{202112CSM,202108MAES}. Unmanned aerial vehicles (UAVs) are becoming increasingly popular as a means of expanding the connectivity and capacity of the existing terrestrial networks~\cite{2022EATNSM,2022TWC,2021Globecom,2022ictc}. For UAV mobile base stations (BSs), a reliable wireless backhaul link is essential to connect UAVs to the core network through a ground gateway. Wireless backhaul systems can be effectively supported by free-space optical (FSO) communications, a promising solution for assisting BSs in NTNs.

FSO communications have distinct characteristics compared to radio frequency (RF)-based wireless communications~\cite{AdvancedFSO}. They provide long-distance communication links but only support point-to-point connections. The transceivers are highly secure against eavesdropping due to the narrow beamwidth, and the high carrier frequency, in the hundreds of terahertz, offers significant bandwidth in an unlicensed band. Because of these advantages, FSO communications are considered suitable for supporting the connection between radio access networks (or a gateway) and mobile BSs across the various non-terrestrial layers of the NTN~\cite{201801ComMag,201806WC,202110JSAC}. For instance, relay UAVs can provide improved data rates to terrestrial FSO networks while ensuring line-of-sight (LoS) conditions~\cite{201801ComMag}. However, one of the primary challenges in establishing an FSO backhaul link is overcoming pointing disturbances caused by atmospheric effects and mechanical jitters. To ensure signal power at the receiver end when transmitting the narrow optical beam, transceivers at both ends must precisely align with each other. This emphasizes the critical role of the pointing, acquisition, and tracking (PAT) system, which manages link establishment, stabilization, and maintenance~\cite{wcl,tvt}.

The presence of a PAT system does not guarantee the elimination of pointing errors. Pointing losses still occur due to residual jitter in the beam footprint at the receiver plane~\cite{2020wcnc,2023EAComMag}. The distribution of pointing errors varies depending on the type of disturbance and can be modeled using different distributions. Without considering boresight error, the pointing error distribution is Rayleigh for one degree of freedom (DoF) and Hoyt for two DoFs~\cite{2015ICC}. Especially, the Rayleigh distribution is widely applicable because it generalizes the sum of all jitter factors into two-dimensional (2D) Gaussian disturbances~\cite{200707JLT}. On the other hand, the Hoyt distribution is suitable for modeling pointing errors due to rotary-wing drone jitters. The authors in~\cite{2018ICC} separately analyze UAV jitters in the vertical and horizontal directions to represent the distinctive jitter characteristics of drones.

To the best of our knowledge, this is the first study that investigates the impact of generalized 3D jitter in UAVs and proposes a method for avoiding highly disturbed link directions by adjusting the UAVs' trajectories. Our results indicate that UAVs should avoid directions significantly affected by specific jitter orientations. The proposed optimization method enhances energy efficiency by adjusting trajectories, taking into account both power consumption and the achievable rate.

	\subsection{Related Works}

\begin{comment}
	\subsubsection{Forward FSO Link}	
As proposed and analyzed in the articles~\cite{uavfsosystemdesign}, the frameworks of the UAV networks with FSO backhaul link has been studied for various future applications. The performance of the forward backhaul link has been analyzed with different network topology. The closed-form outage probability and the optimal altitude of the UAV relay for the mixed FSO/RF communications system is derived in~\cite{201902TAES}. An HAP-to-UAV FSO link is assumed in~\cite{2020VTC} as a part of the satellite-air-ground integrated network (SATIN). The authors analyze the effect of the hovering rotary-wing UAVs on communication link availability and performance. Moreover, in the system model of~\cite{202204TAES}, a satellite-to-UAV FSO backhaul link supports the RF access for multiple ground users. The performance of the ground-air-space forward relay system with FSO link~\cite{202212PJ}, and mixed FSO/RF link~\cite{202101TAES} are also studied as a potential topology of the SATIN.
\end{comment}

\begin{comment}
FSO/RF hybrid with aerial FSO backhaul links (forward link)
[201902TAES]: G-A-A-G relay / only A2A link is FSO / closed-form outage probability and optimal altitude
[2020VTC]: HAP-to-UAV FSO link / Analytical expressions of outage probability and outage capacity
[202204TAES]: SAT-to-UAV FSO link / 

ground-to-air/space, air-to-space
[202101TAES]: uplink
[202212PJ]: UAV-to-SAT uplink backhaul
\end{comment}

	\subsubsection{FSO and Mixed FSO/RF Links}	
	
In our system, the UAV can either be a mobile BS or a source node, and data must be transmitted from the UAV to the ground station (GS) through the reverse link. The performance of the reverse FSO backhaul link has been analyzed and optimized in several studies~\cite{201605TWC,202104TCOM,202204IOTJ}. In~\cite{201605TWC}, the relay node collects data from multiple users through RF access links and then delivers the message to the destination node through a parallel FSO/RF link. Another study explores multi-user data collection at the satellite destination node in the satellite-air-terrestrial integrated network through HAP relay nodes~\cite{202104TCOM}. In~\cite{202204IOTJ}, the authors analyze the mixed FSO/RF relay communications system and derive various performance metrics and analytical features. Unlike other studies, the FSO access link from the UAV to ground users is assumed in~\cite{202208TWC}. Additionally, the cost-efficient communication between the GS and UAVs using a modulating retro-reflector is a promising technology for communicating with UAVs from a fixed GS~\cite{AdvancedFSO,201808TCOM,202208TWC,202111JSAC}.

\begin{comment}
reverse link
[201605TWC]: uplink / source-relay multiple RF access / relay-destination FSO / Multiple users are trying to access to relay / relay collects data and transmits to the destination by FSO link / link association for delay-limited and delay-tolerant condition
[202104TCOM]: Uplink massive access / Ground-to-UAV RF / UAV-to-SAT FSO
[202204IOTJ]: Ground source-to-UAV via RF ($\kappa$-$\mu$ fading) / UAV-to-destination via FSO (Malaga fading) / AF relay / CDF, PDF, MGF, Outage Probability, BER, Ergodic capacity

access link from UAV to ground users
[202208TWC]: SAT-to-UAV via RF (SR fading) / UAV-to-ground via FSO (Malaga fading) / AF relay / CDF, PDF, MGF, Outage Probability, BER, Ergodic capacity

reverse via MRR
MRR based communication between the ground node and the UAV
[201808TCOM]: Full duplex MRR / Gamma-gamma fading
[202208TWC]: uplink and downlink via MRR
[202111JSAC]: Multi-UAV trajectory optimization / MRR
\end{comment}

	\subsubsection{Pointing Error}

FSO communication channels are typically modeled as a product of fast channel fluctuation, atmospheric loss, and pointing loss~\cite{200707JLT}. Particularly for UAV-to-GS links, the transmitter on the UAV becomes more susceptible to pointing errors due to mechanical jitter and payload limitations~\cite{mymag}. The simplest model for pointing error in the context of a Gaussian beam involves a 2D independent and identically distributed (iid) Gaussian error. However, this simple model neglects boresight error and biased jittering. Both of these factors are considered and generalized into five different models in~\cite{2015ICC,201609TWC}, which are applicable to various system models, including fixed-point, shipborne, and UAV-based networks. The distribution of the FSO channel and performance metrics were derived with different pointing error distributions in~\cite{202012JCN} for weak turbulence conditions and in~\cite{202010TCOM} for strong turbulence conditions. The effect of UAV jittering on the pointing error has also been studied for mmWave communications~\cite{202004TWC,202205TWC}. In~\cite{202004TWC}, the authors analyze the effect of vibrations from UAV transceivers on the mmWave channel model. In another study~\cite{202205TWC}, the authors explore the impact of roll, pitch, and yaw jitter of the UAV on the channel training of a 2D antenna array launched at the UAV. The authors transform the 3D jittering model of drones into the jittering of the antenna array in the $x$ and $y$ axes.

\begin{comment}
Pointing error
[200707JLT]: Widely used 2d iid Gaussian
[2015ICC]: Five different pointing error models / boresight and identicalness of vertical and horizontal pointing error
[201609TWC]: expansion of 2015ICC
[202010TCOM]: pointing error models and performance analysis with strong turbulence
[202012JCN]: pointing error models and performance analysis with weak turbulence

[202004TWC]: Effect of UAV fluctuation on RF beam jittering
[202205TWC]: Euler angle jittering of RF antenna array
\end{comment}

	\subsubsection{UAV Trajectory Optimization}	
	
Trajectory optimization has been widely studied as an effective solution for energy-efficient communications using unmanned aerial vehicles (UAVs). Research has been conducted on both RF communications~\cite{201706TWC,201904TWC,201807TVT,202003TVT,201803TWC,202209TWC,202002WC,202011TCOM} and FSO communications~\cite{201801ComMag,202003TWC,202101TWC,202111Entropy,202111JSAC,2022EATVT}. In~\cite{201706TWC} and~\cite{201904TWC}, the energy consumption models for rotary-wing and fixed-wing communication UAVs were proposed, respectively. In both studies, the communication efficiency was optimized using the successive convex approximation (SCA) method to solve for the flight trajectory. The SCA method was used in~\cite{201706TWC,201904TWC,201807TVT,202003TVT} as an optimization technique for finding effective trajectories from non-convex problems. In contrast, the block coordinate descent method was used in~\cite{201803TWC,202209TWC}, and online adaptation through deep reinforcement learning was used in~\cite{202002WC,202011TCOM}. For FSO communication systems, researchers have proposed various flight optimization methods with different criteria. In~\cite{202003TWC}, the link connection time between the GS and the UAV was maximized while considering speed and acceleration constraints. In~\cite{202101TWC,202111Entropy}, the throughput of buffer-aided mixed FSO/RF relay communications was maximized under flight constraints and queueing mechanisms. More complex constraints and objectives were defined in~\cite{202111JSAC,2022EATVT} to optimize the multi-UAV system, taking into account the power consumption of rotary-wing UAVs~\cite{201706TWC}.

\begin{comment}
Energy consumption models of UAV flight
[201706TWC]: Rotary-wing
[201904TWC]: Fixed-wing

non-FSO trajectory optimization

[201807TVT]: high citation SCA
[202003TVT]: SCA / fixed-wing

[201804TWC]: Block coordinate descent (BCD) algorithm
[202209TWC]: BCD algorithm

[202002WC]: DRL
[202011TCOM]: DRL

FSO Trajectory optimization
[201801ComMag]: Importance of FSO trajectory optimization

[2019GLOBECOM]: J.-H. Lee
[202003TWC]: J.-H. Lee / Service time maximization
[202101TWC]: J.-H. Lee / Throughput of buffer-aided mixed FSO/RF relay
[2022EATVT]: Deep reinforcement learning / J.-H. Lee / Intelligent multi-UAV trajectory optimization

[202111Entropy]: Throughput of buffer-aided mixed FSO/RF relay

[202111JSAC]: S. Song / Intelligent multi-UAV trajectory optimization

\end{comment}

	\subsection{Contributions and Paper Organization}

In light of the previous studies, we introduce a new pointing error model for fixed-wing UAVs that accounts for the 3D jitter of UAVs. Based on this model, we present a trajectory optimization algorithm for UAV-assisted FSO communications. Our results indicate that in the selected scenarios, the communication performance is significantly influenced by the position and posture of the UAV relative to the GS. As a result, it is crucial to minimize the impact of the UAV jitter by selecting the least affected flight path. In summary, the key contributions of this work are as follows:

\begin{itemize}

\item{In the context of the UAV-to-GS FSO link, we define the jittering of UAVs in the yaw, pitch, and roll axes. Using this jittering model, we derive the pointing error angle as a Hoyt-distributed random variable. The distribution parameters are determined by the position and posture of the UAV relative to the GS. We formulate the energy efficiency as the ratio of total link capacity to power consumption during the mission flight. The link capacity can be obtained from the link budget of the UAV-to-GS FSO link, which includes the proposed pointing error model. The flight power consumption and kinetic characteristics of the UAVs are obtained from the flight model of the fixed-wing UAVs.}
\item{Using the capacity and power consumption models, we formulate an optimization problem that maximizes the energy efficiency under the speed, acceleration, initial and final coordinates, and elevation angle constraints. Finding the optimal trajectory is a non-trivial problem, primarily due to the highly nonlinear relationship between the acceleration, posture, and link capacity. We utilize the SCA method to convert the non-convex problem into a locally-approximated concave-convex fractional program. During the iteration, the Dinkelbach method is used to maximize the approximated fractional objective function under convex constraints.}
\item{The results of the trajectory optimization problem explicitly show the impact of the generalized 3D jitter of UAVs on the optimal trajectory. Based on these results, we assert that the proposed jittering model should be considered to measure the realistic performance of the FSO link.}

\end{itemize}

The remainder of this work is organized as follows. Section~\RomanNumeralCaps{2} introduces the UAV flight model, FSO link capacity, and 3D jittering of UAVs with mathematical expressions. In Section~\RomanNumeralCaps{3}, we develop the random jittering of fixed-wing UAVs into the pointing error model. The formulation of the energy-efficiency maximization problem is described in Section~\RomanNumeralCaps{4}. The approximation of the non-convex objective function and constraints through the selected optimization techniques is also provided in Section~\RomanNumeralCaps{4}, with theorems, lemmas, and derivations. The numerical results of the generalized pointing error model and the optimization problem are shown in Section~\RomanNumeralCaps{5}. Finally, the paper is concluded in Section~\RomanNumeralCaps{6} with meaningful insights.

	\section{System Model}
	\label{systemmodel}
	
\begin{figure}[t]
	\begin{center}
		{\includegraphics[width=0.95\columnwidth,keepaspectratio]
			{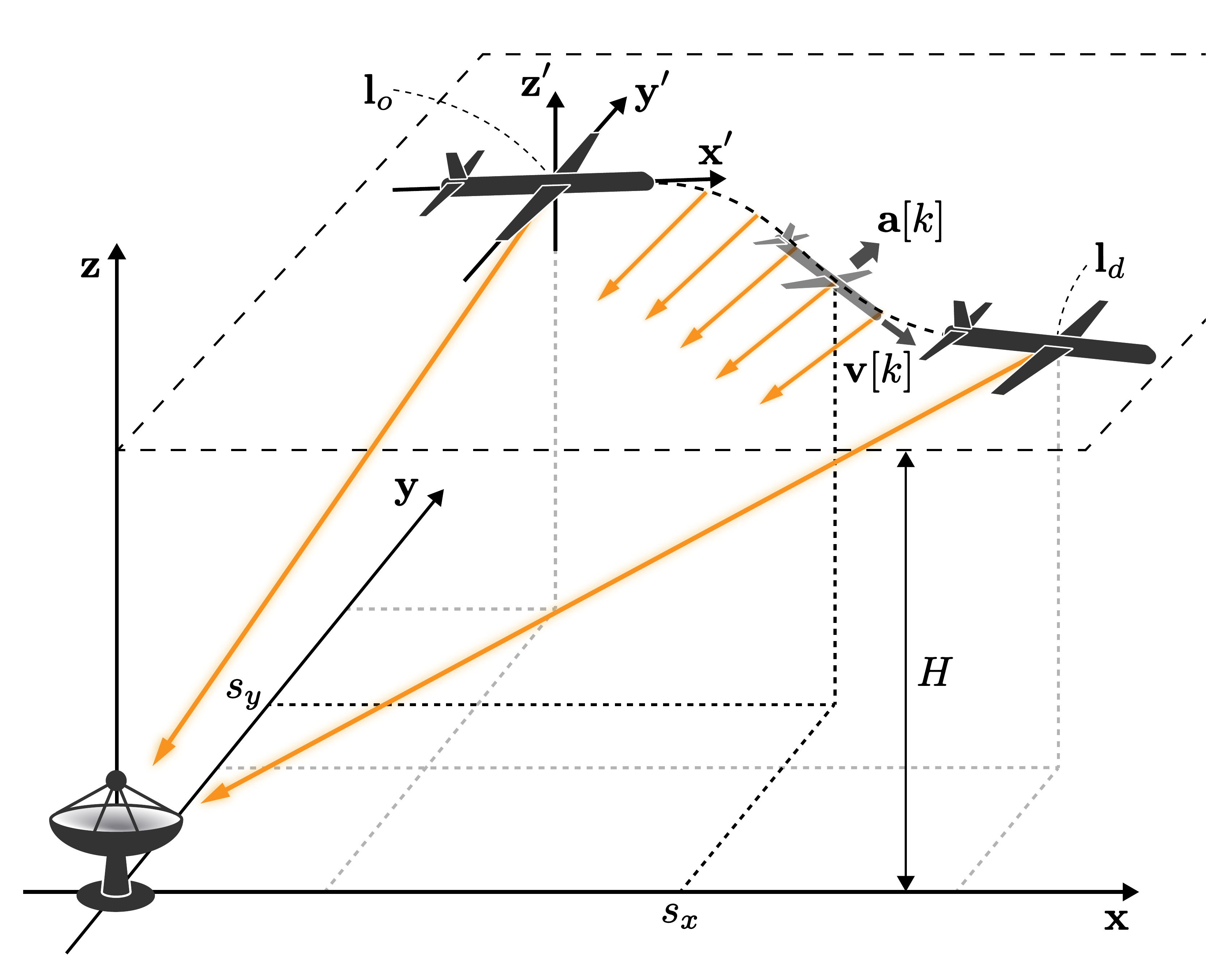}%
			\caption{GS-centered coordinate system.}
			\label{traj01}
		}
	\end{center}
	\vspace{-10pt}
\end{figure}

In this section, we assume that a UAV is performing a mission flight while maintaining an FSO communication link with a fixed GS. The mission flight is subject to constraints with a fixed operating altitude and specified initial and final points. We consider one-hop UAV-to-GS transmission using intensity modulation and direct detection (IM/DD) FSO communications. In order to ensure LoS transmission, the elevation angle of the UAV from the GS is constrained to a certain lower bound. A detailed model of the FSO communication-based mission flight is described in the following subsections. The parameter notations used throughout this paper are organized in Table~\ref{tbl3}.

	\subsection{Discrete-Time Coordinate System}
	\label{distime}
	
To establish a GS-centered coordinate system, as depicted in Fig.~\ref{traj01}, we first discretize continuous time into a small time interval $\delta$ and time slot $k$, where $k\in\mathbb{N}$. The 3D location of the UAV at a given time slot $k$ is expressed in the coordinate system as $\bold{s}[k]$, with the $x$-$y$ plane and $z$ axis corresponding to the ground and altitude, respectively. The posture of the UAV can be defined by the UAV-centered coordinate system, which is represented by $x'$, $y'$, and $z'$ axes. At a given time slot $k$, the heading direction of the UAV and the direction pointed by the left wing are defined as the $x'$ and $y'$ axes, respectively.

We define the time of interest as $k\in[1,N]$, where the mission starts at time slot $1$ and ends at $N$. The velocity and acceleration of the UAV are determined by its position $\bold{s}[k]$. Specifically, the velocity is given by
\begin{equation}
\label{velocity}
\begin{aligned}
&\bold{v}[k]=\frac{\bold{s}[k+1]-\bold{s}[k]}{\delta},\quad1\leq{k}\leq{N-1}.
\end{aligned}
\end{equation}
To facilitate our analysis, we extend the definition of the velocity to time slot $N$ as
\begin{equation}
\label{velocityout}
\begin{aligned}
&\bold{v}[N]=\bold{v}[N-1],
\end{aligned}
\end{equation}
or, equivalently, $\bold{s}[N+1]=2\bold{s}[N]-\bold{s}[N-1]$. The acceleration can be computed as
\begin{equation}
\label{acceleration}
\begin{aligned}
\bold{a}[k]=\frac{\bold{v}[k+1]-\bold{v}[k]}{\delta},\quad1\leq{k}\leq{N-1}.
\end{aligned}
\end{equation}
Throughout this paper, we assume a fixed altitude of the UAV as follows:
\begin{equation}
\label{altitude}
\begin{aligned}
s_z=H,\quad1\leq{k}\leq{N},
\end{aligned}
\end{equation}
where $\bold{s}[k]=[s_x,s_y,s_z]^T$.

{\renewcommand{\arraystretch}{1.05}
\begin{table}[!t]
	\centering
	\caption{Parameters}
	\footnotesize
	\label{tbl3}
	\begin{tabular}{|c|c|}
	\hline
	\bfseries{Parameter} 				& \bfseries{Description} \\ 
	\hhline{|=|=|}
	$x$, $y$, $z$	& Axes of the GS-centered coordinate system\\
	\hline
	$x'$, $y'$, $z'$	& Axes of the UAV-centered coordinate system\\
	\hline
	$\bold{s}[k]$, $\bold{v}[k]$, $\bold{a}[k]$	& Position, velocity, and acceleration at time slot $k$\\
	\hline
	$N$	& Number of time slots\\
	\hline
	$\delta$	& Time interval between slots\\
	\hline
	$\phi$, $\psi$, $\theta$	& Roll, pitch, and yaw angles of the UAV posture\\
	\hline
	$\hat{\bold{u}}[k]$	& Average UAV-to-GS pointing vector\\
	\hline
	$\bold{u}[k]$	& Instantaneous UAV-to-GS pointing vector\\
	\hline
	$C_\mathbb{E}$	& Ergodic capacity of UAV-to-GS transmission\\
	\hline
	$P_\text{T}$	& Transmit power of the UAV\\
	\hline
	$P_\text{F}$	& Flight power consumption of the UAV\\
	\hline
	$E_\text{cost}$	& Launch cost of the flight mission\\
	\hline
	$h_a$, $h_\ell$, $h_p$	& Fast fading, atmospheric loss, and pointing loss\\
	\hline
	$\theta_p$	& UAV-to-GS pointing error angle\\
	\hline
	$\alpha, \beta, \gamma$	& Roll, pitch, and yaw angle jittering of the UAV\\
	\hline
	$v_\text{min}$, $v_\text{max}$	& Minimum and maximum speed of the UAV\\
	\hline
	$a_\text{max}$	& Maximum acceleration of the UAV\\
	\hline
	$\bold{l}_o$, $\bold{l}_d$	& Initial and final position of the mission\\
	\hline
	$\eta$	& Elevation angle of the UAV\\
	\hline
	\end{tabular} 
	\vspace{-15pt}
\end{table}
}

\begin{figure}[t]
	\begin{center}
		{\includegraphics[width=0.95\columnwidth,keepaspectratio]
			{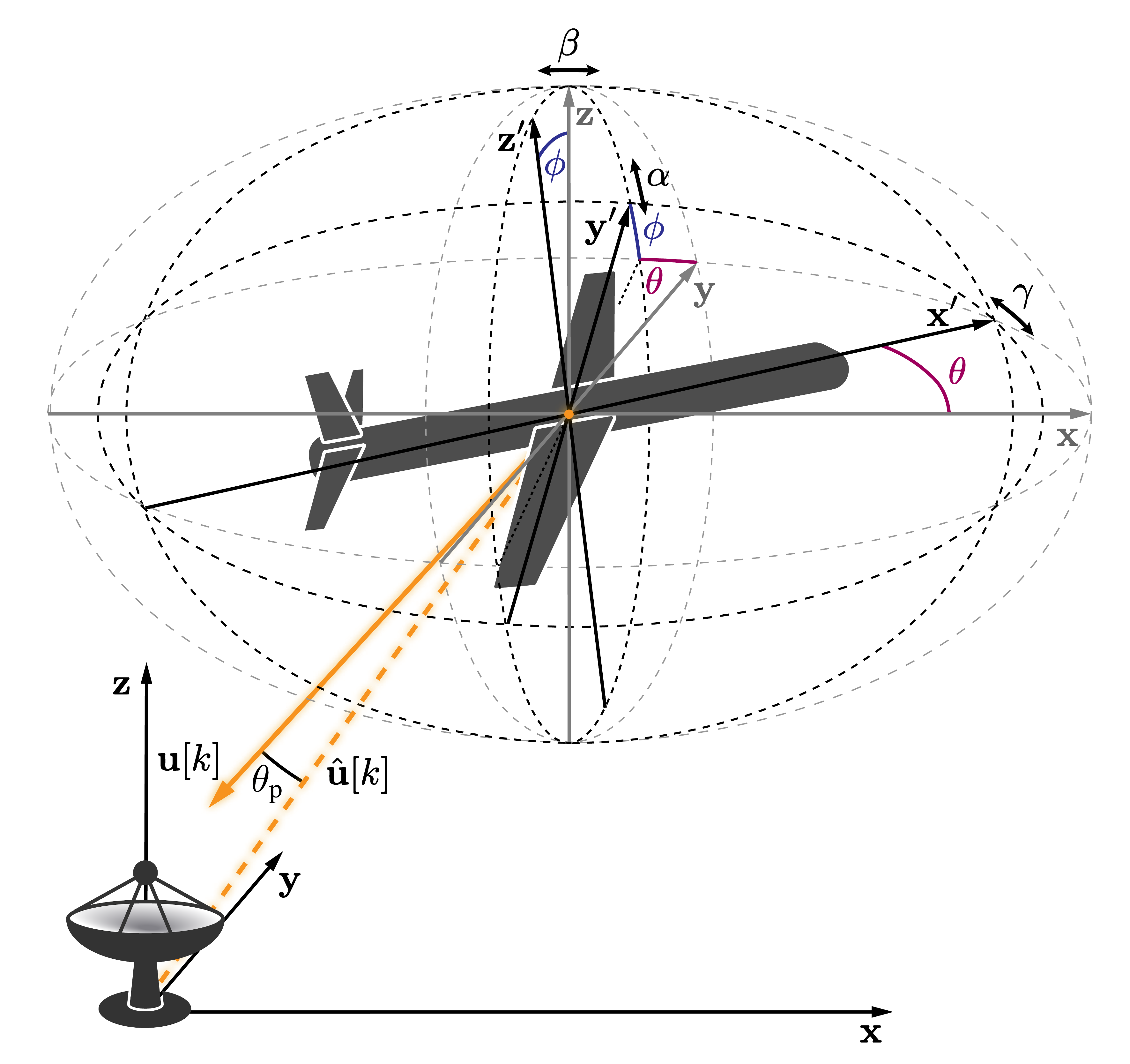}%
			\caption{3D jittering of the UAV in the UAV-centered coordinate system.}
			\label{traj02}
		}
	\end{center}
	\vspace{-10pt}
\end{figure}

To calculate the UAV-to-GS pointing vector in the UAV-centered coordinate system, we need to establish the relationship between the two coordinate systems. The roll, pitch, and yaw angles of the UAV in the GS-centered coordinate system are denoted as $\phi$, $\psi$, and $\theta$, respectively (see Fig.~\ref{traj02}). Assuming a fixed altitude of the UAV, the pitch angle is always zero~\cite{201706TWC}. The yaw angle can be determined using the direction vector of $\bold{v}[k]$, which represents the $x'$ axis in the UAV-centered coordinate system. Specifically, we have
\begin{equation}
\label{uavyaw}
\begin{aligned}
\cos\theta=\frac{v_x}{\norm{\bold{v}[k]}},\,\,\sin\theta=\frac{v_y}{\norm{\bold{v}[k]}},
\end{aligned}
\end{equation}
where $\bold{v}[k]=[v_x,v_y,v_z]^T$. According to~\cite{201706TWC}, acceleration determines the roll angle (or bank angle) of fixed-wing UAVs. Using the formula $\tan\phi=-{a_\perp}/{\textsl{g}}$, we can derive
\begin{equation}
\label{bank}
\begin{aligned}
\tan\phi=\frac{1}{\norm{\bold{v}[k]}\textsl{g}}({v_y}{a_x}-{v_x}{a_y}),
\end{aligned}
\end{equation}
where $a_\perp$ and $\textsl{g}$ are the acceleration in the $y'$ axis and gravitational acceleration, respectively, and $\bold{a}[k]=[a_x,a_y,a_z]^T$. We can express the UAV-to-GS pointing vector, $\hat{\bold{u}}[k]$, in the UAV-centered coordinate system, using the position and posture of the UAV in the GS-centered coordinate system. The 3D rotation matrices with respect to the $x$, $y$, and $z$ axes can be expressed as follows~\cite{202205TWC}:
\begin{equation}
\label{rotmat}
\begin{aligned}
\bold{R}_\text{x}(\alpha)=
\begin{bmatrix}
1	& 0	& 0\\
0	& \cos\alpha	& -\sin\alpha\\
0	& \sin\alpha	& \cos\alpha
\end{bmatrix},\\
\bold{R}_\text{y}(\beta)=
\begin{bmatrix}
\cos\beta	& 0	& \sin\beta\\
0	& 1	& 0\\
-\sin\beta	& 0	& \cos\beta
\end{bmatrix},\\
\bold{R}_\text{z}(\gamma)=
\begin{bmatrix}
\cos\gamma	& -\sin\gamma	& 0\\
\sin\gamma	& \cos\gamma	& 0\\
0	& 0	& 1
\end{bmatrix}.
\end{aligned}
\end{equation}
We then represent $\hat{\bold{u}}[k]$ as
\begin{equation}
\label{vecr}
\begin{aligned}
\hat{\bold{u}}[k]=&-\bold{R}_\text{post}(\phi,\psi,\theta)\bold{s}[k],
\end{aligned}
\end{equation}
where $\bold{R}_\text{post}(\phi,\psi,\theta)=\bold{R}_\text{x}(-\phi)\bold{R}_\text{y}(-\psi)\bold{R}_\text{z}(-\theta)$.
%As described in this section, the introduced variables, $\bold{s}$, $\bold{v}$, $\bold{a}$, $\phi$, $\psi$, and $\theta$, can be used to calculate the flight of the UAV and the link direction in each coordinate system.

	\subsection{Capacity of an FSO Link}
	
The received signal at the GS for IM/DD FSO communications is modeled as in~\cite{202001OJCS}:
\begin{equation}
\label{imddsignal}
\begin{aligned}
\bar{y}=h\bar{x}+\bar{n},
\end{aligned}
\end{equation}
where $\bar{y}$ is the received signal, $h$ is the channel coefficient, $\bar{x}\in\mathbb{R}^{+}$ is the intensity modulated signal, and $\bar{n}$ is the background Gaussian shot noise. The channel $h$ can be modeled as a product of independent factors that account for fast channel fluctuations, atmospheric loss, pointing loss, and responsivity. This is expressed as $h=h_ah_\ell{h_p}R$. The average transmit power is constrained to $P_\text{T}$, and the shot noise follows $\bar{n}\sim\mathcal{N}(0,\sigma^2)$. Then, the average received power at the GS is given by~\cite{200707JLT}
\begin{equation}
\label{rxpower}
\begin{aligned}
P=h_ah_\ell{h_p}RP_\text{T}.
\end{aligned}
\end{equation}
The average optical signal-to-noise ratio (ASNR) for this signal model is $P^2/\sigma^2$. The fast fluctuations of the FSO signal power are mainly caused by scintillation during atmospheric propagation~\cite{AdvancedFSO}. We assume that the channel fluctuation follows a log-normal distribution, given by
\begin{equation}
\label{fading}
\begin{aligned}
f_{h_a}(x)=\frac{1}{2x\sqrt{2\pi\sigma_I^2}}\exp\left(-\frac{(\log{x}+2\sigma_I^2)^2}{8\sigma_I^2}\right),
\end{aligned}
\end{equation}
where $\sigma_I^2$ is the log-amplitude variance~\cite{wcl}. The atmospheric loss follows the Beer-Lambert law as described in
\begin{equation}
\label{loss}
\begin{aligned}
h_\ell=\exp(-\sigma_\text{B}{z}),
\end{aligned}
\end{equation}
 where $\sigma_\text{B}$ is the attenuation coefficient and $z$ represents the propagation length, which is the distance between the UAV and the GS. The attenuation coefficient, given by
\begin{equation}
\label{krusemod}
\begin{aligned}
\sigma_\text{B}=\frac{3.91}{V}\left(\frac{\lambda_\text{FSO}}{\lambda_0}\right)^{-q_\text{sca}},
\end{aligned}
\end{equation}
 depends on the size distribution of the scattering, represented by $q_\text{sca}$, and the wavelength of the FSO link, $\lambda_\text{FSO}$~\cite{AdvancedFSO}. The value of $q_\text{sca}$ is determined by
\begin{equation}
\label{kruseqval}
q_\text{sca}=\begin{cases}
\begin{aligned}
1.6\quad\qquad&V\geq50\,\text{km}\\
1.3\quad\qquad&6\,\text{km}\leq{V}<50\,\text{km}\\
0.585V^{1/3}\quad&V<6\,\text{km}
\end{aligned}
\end{cases}
\end{equation}
where $V$ is the visibility range and $\lambda_0=550$~nm. Assuming a Gaussian beam with a 1-$\sigma$ divergence angle of $2\sigma_\text{div}$, the pointing loss can be described by the following formula~\cite{200707JLT}:
\begin{equation}
\label{pointingloss}
\begin{aligned}
h_p&=A_0\exp\left(-\frac{\theta_p^2}{2\sigma_\text{div}^2}\right),
\end{aligned}
\end{equation}
where $\theta_p$ represents the pointing error angle and $A_0$ is the maximum gain at the center of the beam. The 1-$\sigma$ diameter of the beam footprint at the receiver end can be expressed as $w_b=2z\theta_p$. Let $a$ denote the aperture diameter of the receiver. Assuming that the beamwidth is significantly larger than the aperture size ($w_b\gg{a}$), the maximum gain of $h_p$ can be approximated as $A_0\approx\frac{a^2}{2z\sigma_\text{div}}$~\cite{wcl}.

Based on the channel characteristics, the capacity of the IM/DD FSO communication link can be expressed as follows~\cite{200910TIT}:
\begin{equation}
\label{capacity}
\begin{aligned}
C=\frac{1}{2}\log_2\left(1+\Gamma\right),
\end{aligned}
\end{equation}
where $\Gamma=\frac{e{P}^2}{2\pi\sigma^2}$. Since the channel fluctuation is much faster than the kinematic changes of UAVs~\cite{201108JOCN}, we select the ergodic capacity as a performance metric. The ergodic capacity of the link is defined as follows:
\begin{equation}
\label{ergodicdef}
\begin{aligned}
C_\mathbb{E}=\mathbb{E}\left[\frac{1}{2}\log_2\left(1+\Gamma\right)\right],
\end{aligned}
\end{equation}
where $\mathbb{E}[\cdot]$ represents the expectation of the function. It is important to note that the term $h_p$ in $\Gamma$ depends on the position and posture of the UAV, which significantly influences the value of $C_\mathbb{E}$.

\begin{figure}[t]
	\begin{center}
		{\includegraphics[width=1\columnwidth,keepaspectratio]
			{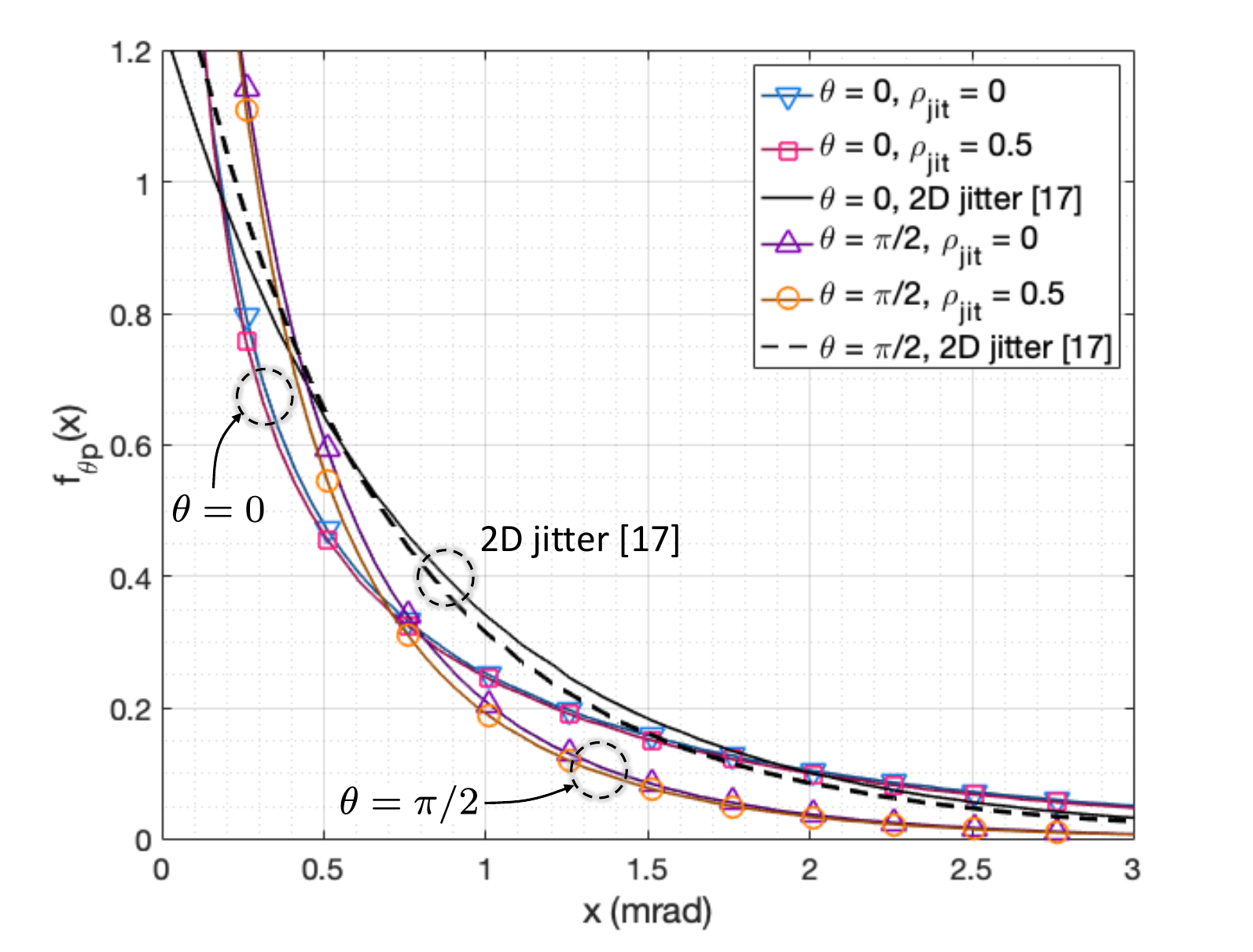}%
			\caption{Probability density function of the pointing error angle.}
			\label{traj03}
		}
	\end{center}
	\vspace{-12pt}
\end{figure}

	\subsection{Flight Power Consumption Model}

A major challenge in UAV communications is power consumption during flight. The power consumption is influenced by the UAV's dynamics, specifically its velocity and acceleration, which are represented by the vectors $\bold{v}$ and $\bold{a}$, respectively. The relationship between these factors and power consumption is given by the following equation~\cite{201706TWC}:
\begin{equation}
\label{flightpowerfw}
\begin{aligned}
P_\text{F}(\bold{v},\bold{a})\approx&{c_1\norm{\bold{v}}^3+\frac{c_2}{\bold{\norm{v}}}\left(1+\frac{\norm{\bold{a}}^2}{\textsl{g}^2}\right)+\frac{\Delta{E_K}}{\delta}},
\end{aligned}
\end{equation}
where $c_1=\frac{1}{2}\rho{C}_{D_0}S$, $c_2=\frac{2W^2}{(\pi{e_0}\mathcal{A}_\mathcal{R})\rho{S}}$,
and $\Delta{E_K}$ is the change in kinetic energy during the time interval of $\delta$. The symbols $\rho$, ${C}_{D_0}$, $S$, $W$, $e_0$, $\mathcal{A}_\mathcal{R}$ denote the air density, zero-lift drag coefficient, reference area, weight, Oswald efficiency, and aspect ratio of the wing, respectively. The total change in kinetic energy can be calculated from the initial and final velocity of the UAV, as follows:
\begin{equation}
\label{pointing0}
\begin{aligned}
\Delta{E}_{K,\text{tot}}=\frac{1}{2}m(\norm{\bold{v}[N]}^2-\norm{\bold{v}[1]}^2),
\end{aligned}
\end{equation}
where $m$ is the mass of the UAV. For a sufficiently long flight time ($\delta{N}\gg{1}$), we can ignore $\Delta{E}_{K,\text{tot}}$, since $\norm{\bold{v}[1]}$ and $\norm{\bold{v}[N]}$ are both restricted by the maximum and minimum velocity constraints. Thus, the power consumption at time slot~$k$ is approximated as follows:
\begin{equation}
\label{totflightpowerfw}
\begin{aligned}
P_\text{F}&(\bold{v}[k],\bold{a}[k])\approx{c_1}\norm{\bold{v}[k]}^3+\frac{c_2}{\norm{\bold{v}[k]}}\left(1+\frac{\norm{\bold{a}[k]}^2}{\textsl{g}^2}\right).
\end{aligned}
\end{equation}
Through trajectory optimization, the power consumption can be economized by determining the most energy-efficient flight path for operating the mission.

	\section{Pointing Error Analysis}

In this section, we introduce the 3D jittering model of the UAV, building upon the system model outlined in Section~\ref{distime}. To describe the instantaneous UAV-to-GS pointing vector, which incorporates the impact of residual jittering in the PAT system, we develop the jittering matrix. We then establish the pointing error angle as the deviation between the instantaneous pointing vector and the average pointing vector, excluding the jittering effect. Finally, we derive the distribution of the pointing error angle.
	
	\subsection{3D Jittering Model of Fixed-wing Aircraft}
	\label{3djit}

By analyzing the flight model and the relationship between the two coordinate systems introduced in Section~\ref{distime}, we obtain $\hat{\bold{u}}[k]$, a UAV-to-GS pointing vector when the posture of the UAV is fixed along the $x'$, $y'$, and $z'$ axes. To account for the effect of the jitter, we first express it as a transformation matrix. From~(\ref{rotmat}), the jittering matrix can be expressed as follows:
\begin{equation}
\label{jitter}
\begin{aligned}
\bold{R}_\text{jitter}(\alpha,\beta,\gamma)=&\bold{R}_\text{x}(\alpha)\bold{R}_\text{y}(\beta)\bold{R}_\text{z}(\gamma)\\
\approx&
\begin{bmatrix}
1	& -\gamma	& \beta\\
\gamma	& 1	& -\alpha\\
-\beta	& \alpha	& 1
\end{bmatrix},
\end{aligned}
\end{equation}
where $\alpha$, $\beta$, and $\gamma$ are roll, pitch, and yaw angle jittering of the UAV, respectively. We apply small angle approximations ($\sin\epsilon\approx\epsilon$ and $\cos\epsilon\approx{1}$) to simplify calculations. We consider the 3D jittering angles as random variables, represented by the vector $\bold{x}=\left[\alpha,\beta,\gamma\right]^T$ that follows a multivariate Gaussian distribution $\bold{x}\sim\mathcal{N}(\bold{\mu},\bold{\Sigma})$. The mean and covariance of $\bold{x}$ are defined by
\begin{equation}
\label{rvvecpara}
\begin{aligned}
\bold{\mu}=[0,0,0]^T,\,\bold{\Sigma}=
\begin{bmatrix}
\sigma_\alpha^2	& \rho_{\alpha,\beta}{\sigma_\alpha}{\sigma_\beta}	& \rho_{\gamma,\alpha}{\sigma_\gamma}{\sigma_\alpha}\\
\rho_{\alpha,\beta}{\sigma_\alpha}{\sigma_\beta}	& \sigma_\beta^2	& \rho_{\beta,\gamma}{\sigma_\beta}{\sigma_\gamma}\\
\rho_{\gamma,\alpha}{\sigma_\gamma}{\sigma_\alpha}	& \rho_{\beta,\gamma}{\sigma_\beta}{\sigma_\gamma}	& \sigma_\gamma^2
\end{bmatrix}.
\end{aligned}
\end{equation}
The covariance matrix $\bold{\Sigma}$ represents the jittering characteristics of the fixed-wing UAV. Now we develop the instantaneous pointing vector, including the effect of the 3D jitter, as
\begin{equation}
\label{pointing1}
\begin{aligned}
\bold{u}[k]=&-\bold{R}_\text{jitter}(\alpha,\beta,\gamma)\bold{R}_\text{post}(\phi,\psi,\theta)\bold{s}[k].
\end{aligned}
\end{equation}
It can be equivalently expressed as $\bold{u}[k]=\bold{R}_\text{jitter}(\alpha,\beta,\gamma)\hat{\bold{u}}[k]$.
Finally, the pointing error angle $\theta_p$ is given by
\begin{equation}
\label{pedef}
\begin{aligned}
\sin\theta_\text{p}=\frac{\lvert\bold{u}[k]\times\hat{\bold{u}}[k]\rvert}{\lvert\bold{u}[k]\rvert\lvert\hat{\bold{u}}[k]\rvert},
\end{aligned}
\end{equation}
which characterizes the angle difference between the average pointing vector $\hat{\bold{u}}[k]$ and the instantaneous pointing vector $\bold{u}[k]$ from the perspective of the UAV.

	\subsection{Distribution of the Pointing Error Angle}
	
When the boresight error is neglected, the pointing error angle is defined as the angle difference between the actual pointing vector $\bold{u}[k]$, which incorporates the effect of the 3D jitter, and the average pointing vector $\hat{\bold{u}}[k]$. We proceed to analyze the characteristics of $\theta_p$, the random variable defined in~(\ref{pedef}).

{\renewcommand{\arraystretch}{1.05}
\begin{table}[!t]
	\centering
	\caption{Comparison of the pointing error models}
	\footnotesize
	\label{tbl1}
	\begin{tabular}{|c|c|c|c|c|}
	\hline
	\bfseries{Model}	& \bfseries{Parameter}		& $\lambda_1$	& $\lambda_2$	& \begin{tabular}{@{}c@{}}$\mathbb{E}[\theta_p^2]=$ \\ $\lambda_1+\lambda_2$\end{tabular}\\
	\hhline{|=|=|=|=|=|}
	Ours	& $\theta=0$, $\rho_\text{jit}=0$	& $0.9664$	& $0.0522$	& $1.0186$\\
	\hline
	Ours	& $\theta=0$, $\rho_\text{jit}=0.5$	& $0.9202$	& $0.0324$	& $0.9526$\\
	\hline
	2D Model	& $\theta=0$~\cite{2018ICC}	& $0.5449$	& $0.2827$	& $0.8276$\\
	\hhline{|=|=|=|=|=|}
	Ours	& $\theta=\pi/2$, $\rho_\text{jit}=0$	& $0.3797$	& $0.0891$	& $0.4688$\\
	\hline
	Ours	& $\theta=\pi/2$, $\rho_\text{jit}=0.5$	& $0.3723$	& $0.0640$	& $0.4363$\\
	\hline
	2D Model	& $\theta=\pi/2$~\cite{2018ICC}	& $0.5449$	& $0.2074$	& $0.7523$\\
	\hline
	\end{tabular} 
	\vspace{-10pt}
\end{table}
}

\newtheorem{theorem}{Theorem}
\begin{theorem}
\label{theo1}
{The distribution of $\theta_p$ follows the Hoyt distribution as
\begin{equation}
\label{pervs}
\begin{aligned}
\theta_p\sim\mathcal{H}(q,\Omega),
\end{aligned}
\end{equation}
where $q=\sqrt{\lambda_1/\lambda_2}$ and $\Omega=\lambda_1+\lambda_2$~\cite{2018ICC}. The symbols $\lambda_1$, $\lambda_2$ ($\lambda_1\geq\lambda_2$) are the nonzero eigenvalues of $\bold{\Sigma}^{1/2}\bold{A}\bold{\Sigma}^{1/2}$, where $\bold{A}$ is defined as
%where $\zeta_1^2$, $\zeta_2^2$, $\zeta_3^2$ are the iid chi-squared random variables as $\zeta_1^2,\zeta_2^2,\zeta_3^2\sim\mathcal{X}^2(1)$ and $\lambda_1$, $\lambda_2$, $\lambda_3$ are the eigenvalues of $\bold{\Sigma}^{1/2}\bold{A}\bold{\Sigma}^{1/2}$. The matrix $\bold{A}$ is defined as
\begin{comment}
\begin{equation}
\label{quadratic}
\begin{aligned}
\bold{A}=\frac{1}{z^2}
\begin{bmatrix}
\hat{{u}}_y^2+\hat{{u}}_z^2	& -\hat{{u}}_x\hat{{u}}_y	& -\hat{{u}}_z\hat{{u}}_x\\
-\hat{{u}}_x\hat{{u}}_y	& \hat{{u}}_z^2+\hat{{u}}_x^2	& -\hat{{u}}_y\hat{{u}}_z\\
-\hat{{u}}_z\hat{{u}}_a	& -\hat{{u}}_y\hat{{u}}_z	& \hat{{u}}_x^2+\hat{{u}}_y^2
\end{bmatrix},
\end{aligned}
\end{equation}
\end{comment}
\begin{equation}
\label{quadraticaa}
\begin{aligned}
\bold{A}=\bold{I}-\frac{1}{z^2}{\hat{\bold{u}}[k]}{\hat{\bold{u}}[k]}^T,
\end{aligned}
\end{equation}
with the identity matrix $\bold{I}$, and $z={({\hat{\bold{u}}[k]}^T{\hat{\bold{u}}[k]})}^{1/2}$ representing the link distance.
}
\end{theorem}

\def\QEDmark{\ensuremath{\blacksquare}}
\def\proof{\emph{Proof: }}
\def\endproof{\hfill\QEDmark}

\proof
See Appendix~\ref{appen1}.
\endproof

During the proof of \emph{Theorem}~\ref{theo1}, we obtain an expression for the expected value of $\theta_p^2$:
\begin{equation}
\label{peexpect}
\begin{aligned}
\mathbb{E}[\theta_p^2]=\lambda_1+\lambda_2,
\end{aligned}
\end{equation}
which can also be computed using $\text{Tr}(\bold{\Sigma}^{1/2}\bold{A}\bold{\Sigma}^{1/2}) = \lambda_1 + \lambda_2$. Fig.~\ref{traj03} presents a numerical comparison of the probability density function of $\theta_p$ for our generalized model, which incorporates the 3D jittering characteristics, and the pointing error model from~\cite{2018ICC}, that considers only two DoFs of drone jitter. In each case illustrated in Fig.~\ref{traj03}, the angles $\psi=-10^\circ$ and $\phi=0$ are held constant. We simulate different values of $\theta$ and the correlation coefficients $\rho_{\alpha,\beta}=\rho_{\beta,\gamma}=\rho_{\gamma,\alpha}=\rho_\text{jit}$. The standard deviations for the jittering along each axis are set to $\sigma_\alpha = 1$ mrad for roll, $\sigma_\beta = 0.3$ mrad for pitch, and $\sigma_\gamma = 0.1$ mrad for yaw. In the 2D jitter model, where $\sigma_\alpha$ and $\sigma_\beta$ are not separately defined, we use $\sigma_\text{x-y} = \sqrt{(\sigma_\alpha^2 + \sigma_\beta^2)/2} = 0.738$ mrad and $\sigma_\gamma = 0.1$ mrad to best represent the actual jitter characteristics using only two DoFs. The positions of the GS and the UAV are assumed to be $[0,0,0]$~m and $[50,550,600]$~m, respectively. We evaluate the pointing error angle under two different heading directions, $\theta=0$ and $\theta=\pi/2$. The results depicted in Fig.~\ref{traj03} and Table~\ref{tbl1} highlight significant differences in the distribution between the proposed generalized pointing error model and the model proposed in~\cite{2018ICC}. It emphasizes the importance of considering the UAV-specific 3D jittering characteristics in link performance assessments.

\begin{comment}
\begin{equation}
\label{exbeamdis}
\begin{aligned}
\mathbb{E}[d_\text{p}^2]=\mathrm{Tr}(\bold{\Sigma}^{1/2}\bold{A}\bold{\Sigma}^{1/2})=z^2(\lambda_1+\lambda_2+\lambda_3)
\end{aligned}
\end{equation}

Beam displacement
\begin{equation}
\label{beamdis}
\begin{aligned}
d_\text{p}^2=z^2(\lambda_1\zeta_1^2+\lambda_2\zeta_2^2+\lambda_3\zeta_3^2)
\end{aligned}
\end{equation}

Pointing error angle

\end{comment}

	\section{Trajectory Optimization for Mission Flight}
	
In this section, we first define the trajectory optimization problem aimed at maximizing energy efficiency. Since the problem is highly non-convex, we iteratively solve using the SCA method. Each iteration of the SCA method involves approximating the non-convex functions, and we provide detailed descriptions of these approximations. Subsequently, we utilize the Dinkelbach method to convert the problem, which originally has a fractional objective function, into an equivalent convex problem that can be addressed iteratively.

	\subsection{Problem Formulation}
	
Based on the system model introduced in Section~\ref{systemmodel}, we aim to optimize the trajectory of the UAV over time to maximize the energy efficiency. Throughout this section, we determine $\bold{s}[k]$, the position of the UAV in the GS-centered coordinate system at each time slot $1\leq{k}\leq{N}$. We formulate this problem as follows:
\begin{subequations}
\label{problem}
\begin{align}
\text{(P1)}:\,\,
\begin{split}
\label{po}
\max_{\bold{s}[k]}\quad&\frac{\sum_{k=1}^{N}C_\mathbb{E}[k]}{\sum_{k=1}^{N-1}P_\text{F}(\bold{v}[k],\bold{a}[k])+NP_\text{T}+{E_\text{cost}}/{\delta}}
\end{split}
\\
\begin{split}
\label{pc1}
\textrm{s.t.}\quad&v_\text{min}\leq\norm{\bold{v}[k]}\leq{{v}_\text{max}},\quad1\leq{k}\leq{N}
\end{split}
\\
\begin{split}
\label{pc2}
&\norm{\bold{a}[k]}\leq{a_\text{max}},\quad1\leq{k}\leq{N-1}
\end{split}
\\
\begin{split}
\label{pc00}
&(\ref{velocity}),(\ref{velocityout}),(\ref{acceleration}),(\ref{altitude})
\end{split}
\\
\begin{split}
\label{pc3}
&\bold{s}[1]=\bold{l}_o,\,\bold{s}[N]=\bold{l}_d
\end{split}
\\
\begin{split}
\label{pc5}
&\frac{\pi}{4}\leq\eta\leq\frac{\pi}{2},{\quad}1\leq{k}\leq{N}
\end{split}
\end{align}
\end{subequations}
where $E_\text{cost}$ is the launch cost of the mission, $v_\text{min}$ and $v_\text{max}$ represent the maximum and minimum speed, respectively, and $a_\text{max}$ is the maximum acceleration. The parameters $\bold{l}_o$ and $\bold{l}_d$ represent the initial and final position, respectively. The elevation angle $\eta$ is defined as $\eta=\arctan({s_z}/{\sqrt{s_x^2+s_y^2}})$ where $\bold{s}[k]=[s_x,s_y,s_z]^T$. The constraints in (P1) include (\ref{pc1}) for the minimum and maximum speed of the UAV, (\ref{pc2}) for the maximum acceleration, (\ref{pc00}) for the relationship between position, velocity, and acceleration, (\ref{pc3}) for the initial and final points of the mission. We also assume the elevation angle constraint in (\ref{pc5}), which guarantees the LoS condition. While the constraints (\ref{pc00}), (\ref{pc3}), and (\ref{pc5}) are convex, (\ref{po}), (\ref{pc1}), and (\ref{pc2}) are non-convex. To solve this highly non-convex problem, we utilize the SCA method.

	\subsection{Successive Convex Approximation}
	\label{scasection}
	
In this subsection, we first approximate the non-convex objective function~(\ref{po}) to a concave-convex fractional function. Specifically, we approximate the numerator to a concave function using the logarithmic approximation and by introducing auxiliary variables, while the denominator is approximated to a convex function by introducing an auxiliary variable. Then,~(\ref{pc1}) and (\ref{pc2}) are approximated to convex constraints. During the SCA method, we utilize the optimized variables from the $p$-th iteration as known parameters in the $(p+1)$-th iteration.
	
To convert~(\ref{capacity}) into a more tractable function, we apply logarithmic approximation, which provides a tight lower bound of the original function~\cite{201710JSAC}. The parameters of the logarithmic approximation can be obtained during the iterations of the SCA method. We can approximate the capacity function for the $(p+1)$-th iteration utilizing the optimization results of the $p$-th iteration. This approximation offers the lower bound as follows:
\begin{equation}
\label{logarithmicapprox}
\begin{aligned}
\log_2\left(1+\Gamma\right)\geq\frac{1}{\log2}(\nabla_\text{L}^{(p)}\log\Gamma+\delta_\text{L}^{(p)}),
\end{aligned}
\end{equation}
where the parameters can be obtained using $\Gamma_\text{L}^{(p)}$ for the tightness at $\Gamma=\Gamma_\text{L}^{(p)}$ as
\begin{equation}
\label{logarithmicparam}
\begin{aligned}
\nabla_\text{L}^{(p)}=\frac{\Gamma_\text{L}^{(p)}}{1+\Gamma_\text{L}^{(p)}},\,\delta_\text{L}^{(p)}=\log(1+\Gamma_\text{L}^{(p)})-\frac{\Gamma_\text{L}^{(p)}}{1+\Gamma_\text{L}^{(p)}}\log\Gamma_\text{L}^{(p)}.
\end{aligned}
\end{equation}
We introduce a new variable $\Gamma_\text{L}^{(p)}$ to avoid confusion with $\Gamma^{(p)}$, which can be interpreted as a random value of $\Gamma$ for the given variables in the $p$-th iteration.
From~(\ref{capacity}) and~(\ref{logarithmicapprox}), the approximated ergodic capacity during the iteration of the SCA method is derived as follows:
\begin{equation}
\label{finalergodic}
\begin{aligned}
C_\mathbb{E}=\frac{1}{2\log2}(\nabla_\text{L}^{(p)}\mathbb{E}[\log\Gamma]+\delta_\text{L}^{(p)}).
\end{aligned}
\end{equation}
\newtheorem{lemma}{Lemma}
\begin{lemma}
\label{theo2}
Using $\Gamma=\frac{e{P}^2}{2\pi\sigma^2}$, we can express $\mathbb{E}[\log\Gamma]$ in~(\ref{finalergodic}) as follows:
\begin{equation}
\label{capacitytheo}
\begin{aligned}
\mathbb{E}[\log\Gamma]=&c_3-2\sigma_\text{B}{z}-2\log{z}-\frac{\lambda_1+\lambda_2}{z^2\sigma_\text{div}^2},
\end{aligned}
\end{equation}
where $c_3=\log\left(\frac{e{R}^2{P_\text{T}}^2a^4}{8\pi\sigma^2\sigma_\text{div}^2}\right)-4\sigma_I^2$ and $z$, $\lambda_1$, $\lambda_2$ are the functions of $\hat{\bold{u}}[k]$ which are defined as in \emph{Theorem}~\ref{theo1}.
\end{lemma}

\proof
See Appendix~\ref{appen2}.
\endproof

\begin{algorithm}[t]
	\caption{Algorithm for Solving (P1) Using the SCA and Dinkelbach Methods.}
	\label{alg1}
	\begin{algorithmic}[1]
			\State Let the set of original variables as $\mathbb{V}=\{\bold{s},\hat{\bold{u}},\bold{v},\bold{a},\boldsymbol{\kappa}\}$, and the set of auxiliary variables as $\mathbb{A}=\{S,\check{U},\check{V},\check{P},Q,R\}$.
			\State Let $\mathbb{V}^{(p)}=\{\bold{s}^{(p)},\hat{\bold{u}}^{(p)},\bold{v}^{(p)},\bold{a}^{(p)},\boldsymbol{\kappa}^{(p)}\}$ and $\mathbb{A}^{(p)}=\{S^{(p)},\check{U}^{(p)},\check{V}^{(p)},\check{P}^{(p)},Q^{(p)},R^{(p)}\}$.
			\State For $k\in\mathbb{N},\,1\leq{k}\leq{N}$, initialize $\bold{s}^{(0)}[k]$, $\hat{\bold{u}}^{(0)}[k]$, $\bold{v}^{(0)}[k]$, $\bold{a}^{(0)}[k]$, $\boldsymbol{\kappa}^{(0)}[k]$, $S_k^{(0)}$, $\check{U}_k^{(0)}$, $\check{V}_k^{(0)}$, and for $k\in\mathbb{N},\,1\leq{k}\leq{N-1}$, initialize $\check{P}_k^{(0)}$, $Q_k^{(0)}$, $R_k^{(0)}$.
			\State Set threshold $\tau_v$.
			\State $p=0$.
			\Repeat
				\State $\forall k$, obtain $\Gamma_{\text{L},k}^{(p)}$, $\nabla_{\text{L},k}^{(p)}$, and $\delta_{\text{L},k}^{(p)}$ through (\ref{gammap}) and (\ref{logarithmicparam}).
				\State Set $\lambda_\text{min}$, $\lambda_\text{max}$, and $\lambda^{(0)} = (\lambda_\text{min}+\lambda_\text{max})/2$.
				\State Set threshold $\tau_f$.
				\State $i=0$.
				\Repeat
					\State Solve (P3), which is a convex problem, and find the optimal value, $F$.
					\If{$F>0$}
						\State $\lambda^{(i+1)} = (\lambda^{(i)}+\lambda_\text{min})/2$.
					\Else
						\State $\lambda^{(i+1)} = (\lambda^{(i)}+\lambda_\text{max})/2$.
					\EndIf
					\State $i\gets{i+1}$.
				\Until{$|F|\leq{\tau_f}$.}
			\State For $\mathbb{V}$ and $\mathbb{A}$, the set of the final solution, $\mathbb{V}^{(p+1)}=\mathbb{V}$ and $\mathbb{A}^{(p+1)}=\mathbb{A}$.
			\State $p\gets{p+1}$.
			\Until{$\norm{\mathbb{V}^{(p)}-\mathbb{V}^{(p-1)}}+\norm{\mathbb{A}^{(p)}-\mathbb{A}^{(p-1)}}<\tau_v$.}
			\State \textbf{end}
	\end{algorithmic}
\end{algorithm}

Now that we place the form of $\log_2\left(1+\Gamma\right)$ with the form of $\log\Gamma$, we can directly substitute $(\ref{capacitytheo})$ into $(\ref{finalergodic})$. We apply the above logarithmic approximation and introduce auxiliary variables to transform the numerator and denominator of the objective function into concave and convex functions, respectively. This transformation allows us to apply the Dinkelbach method, assuming that the numerator and denominator of~(\ref{po}) are always positive, which is a natural assumption.
%The transformed objective function is given by~(P2), obtained from~(\ref{finalergodic}),~(\ref{capacitytheo}), and several auxiliary variables with additional constraints.
The superscript~$(p)$ denotes the result parameter value obtained from the $p$-th problem. Note that (P2) is the problem of the $(p+1)$-th iteration of the SCA method, and the objective is to find the $(p+1)$-th optimized variables.
\begin{subequations}
\label{problemsca}
\begin{align}
\text{(P2)}:\notag\\
\begin{split}
\label{obj2}
\max_{\bold{s}[k]}\quad&\frac{\sum_{k=1}^{N}\nabla_{\text{L},k}^{(p)}(c_3-c_4\norm{\bold{s}[k]}-c_5\check{U}_k^2-c_6\check{V}_k)+\delta_{\text{L},k}^{(p)}}{\sum_{k=1}^{N-1}\check{P}_k+NP_\text{T}+{E_\text{cost}}/{\delta}}
\end{split}\\
\begin{split}
\label{vnc1}
\textrm{s.t.}\quad&v_\text{min}\leq\norm{\bold{v}[k]}\leq{{v}_\text{max}},\quad1\leq{k}\leq{N}
\end{split}\\
\begin{split}
\label{convexc}
&\norm{\bold{a}[k]}\leq{a_\text{max}},\quad1\leq{k}\leq{N-1}
\end{split}\\
\begin{split}
\label{convexd}
&(\ref{velocity}),(\ref{velocityout}),(\ref{acceleration}),(\ref{altitude})
\end{split}\\
\begin{split}
\label{nc1}
&S_k^2\leq\norm{\bold{s}[k]}^2,\quad\forall{k}
\end{split}\\
\begin{split}
\label{nc2}
&S_k\check{U}_k\geq\sqrt{{\hat{\bold{u}}[k]}^T\bold{D}\hat{\bold{u}}[k]},\quad\forall{k}
\end{split}\\
\begin{split}
\label{nc3}
&\check{V}_k\geq\log(\norm{\bold{s}[k]}),\quad\forall{k}
\end{split}\\
\begin{split}
\label{nc4}
&\bold{s}[1]=\bold{l}_o,\,\bold{s}[N]=\bold{l}_d
\end{split}\\
\begin{split}
\label{nc5}
&\check{P}_k\geq{c_1}\norm{\bold{v}[k]}^3+c_2Q_k,\quad1\leq{k}\leq{N-1}
\end{split}\\
\begin{split}
\label{nc6}
&R_k^2\leq\norm{\bold{v}[k]}^2,\quad1\leq{k}\leq{N-1}
\end{split}\\
\begin{split}
\label{nc7}
&Q_kR_k\geq1+\frac{\norm{\bold{a}[k]}^2}{\textsl{g}},\quad1\leq{k}\leq{N-1}
\end{split}\\
\begin{split}
\label{elev2}
&\frac{\pi}{4}\leq\eta\leq\frac{\pi}{2},\quad\forall{k}
\end{split}
\end{align}
\end{subequations}
The parameters $\nabla_{\text{L},k}^{(p)}$ and $\delta_{\text{L},k}^{(p)}$ can be obtained using (\ref{logarithmicparam}). We can adjust $\Gamma_{\text{L},k}^{(p)}$ heuristically, by setting it to
\begin{equation}
\label{gammap}
\begin{aligned}
\Gamma_{\text{L},k}^{(p)}=\exp(&c_3-c_4\norm{\bold{s}^{(p)}[k]}-c_5\check{U}_k^{(p)2}-c_6\check{V}^{(p)}),
\end{aligned}
\end{equation}
which is obtained by $\Gamma_{\text{L},k}^{(p)}=e^{\mathbb{E}[\log\Gamma_k^{(p)}]}$, where $\Gamma_k^{(p)}$ is a random value of $\Gamma$ at time slot $k$ in the $p$-th iteration. The symbols $c_4$, $c_5$, and $c_6$ are constants with values of $c_4=2{\sigma_\text{B}}$, $c_5={1}/{\sigma_\text{div}^2}$, and $c_6=2$.

{\renewcommand{\arraystretch}{1.05}
\begin{table}[!t]
	\centering
	\caption{Simulation parameters}
	\small
	\label{tbl2}
	\begin{tabular}{|c|c|}
	\hline
	\bfseries{Parameter} 				& \bfseries{Value} \\ 
	\hhline{|=|=|}
	Altitude ($H$)	& $600\,\,\si{\metre}$\\
	\hline
	Transmit power ($P_\text{T}$)	& $10\,\,\si{\milli\watt}$\\
	\hline
	$P_\text{T}/\sigma$	& $30\,\,\mathrm{dB}$\\
	\hline
	%FWHM beam divergence	& $2\,\,\si{\milli\radian}$\\
	Responsivity ($R$)	& $0.5\,\,\mathrm{A}/\mathrm{W}$\\
	\hline
	Aperture diameter ($a$)	& $20\,\,\si{\centi\meter}$\\
	\hline
	Log-amplitude standard deviation ($\sigma_I$)	& $0.3$\\
	\hline
	Visibility range ($V$)	& $3\,\,\si{\kilo\meter}$\\
	\hline
	Attenuation coefficient ($\sigma_\text{B}$)	& $5.44\times{10^{-4}}\,\,\mathrm{dB}/\mathrm{km}$\\
	\hline
	Time interval between slots ($\delta$)	& $0.2\,\,\si{\second}$\\
	\hline
	Minimum speed ($v_\text{min}$)	& $3\,\,\mathrm{m}/\mathrm{s}$\\
	\hline
	Maximum speed ($v_\text{max}$)	& $100\,\,\mathrm{m}/\mathrm{s}$\\
	\hline
	Maximum acceleration ($a_\text{max}$)	& $5\,\,\mathrm{m}/\mathrm{s}^2$\\
	\hline
	Gravitational acceleration ($g$)	& $9.8\,\,\mathrm{m}/\mathrm{s}^2$\\
	%Nearest CCR spacing ($\rho_\alpha, \rho_\beta$)	& $\sqrt{2}\,\,\si{\metre}$\\
	\hline
	\end{tabular} 
	\vspace{-15pt}
\end{table}
}

In the formulation process of (P2), the total capacity $\sum_{k=1}^{N}C_\mathbb{E}[k]$ in (P1) is approximated by~(\ref{finalergodic}) and~(\ref{capacitytheo}). Also, we introduce the auxiliary variables $\check{U}_k$ and $\check{V}_k$, which are constrained by (\ref{nc2}) and (\ref{nc3}), respectively. Here, (\ref{nc2}) is derived from $\frac{\lambda_1+\lambda_2}{z^2}=\frac{{\hat{\bold{u}}[k]}^T\bold{D}\hat{\bold{u}}[k]}{\norm{\bold{s}[k]}^2}$, where $\bold{D}=\bold{\text{diag}}(\sigma_\beta^2+\sigma_\gamma^2,\sigma_\gamma^2+\sigma_\alpha^2,\sigma_\alpha^2+\sigma_\beta^2)$. Due to the introduced variables, the numerator of~(\ref{obj2}) becomes concave. To make the denominator convex, we introduce $\check{P}_k$ which is constrained by~(\ref{nc5}). Still,~(\ref{vnc1}), (\ref{nc1}), (\ref{nc2}), (\ref{nc3}), (\ref{nc6}), and (\ref{nc7}) are non-convex constraints. Each should be replaced with the approximated convex constraints to apply the SCA method.

\begin{figure*}[t]
	%\begin{center}
	    \centering
\subfloat[\label{line_traj_narr}]{%
       \includegraphics[width=0.62\columnwidth,keepaspectratio]
			{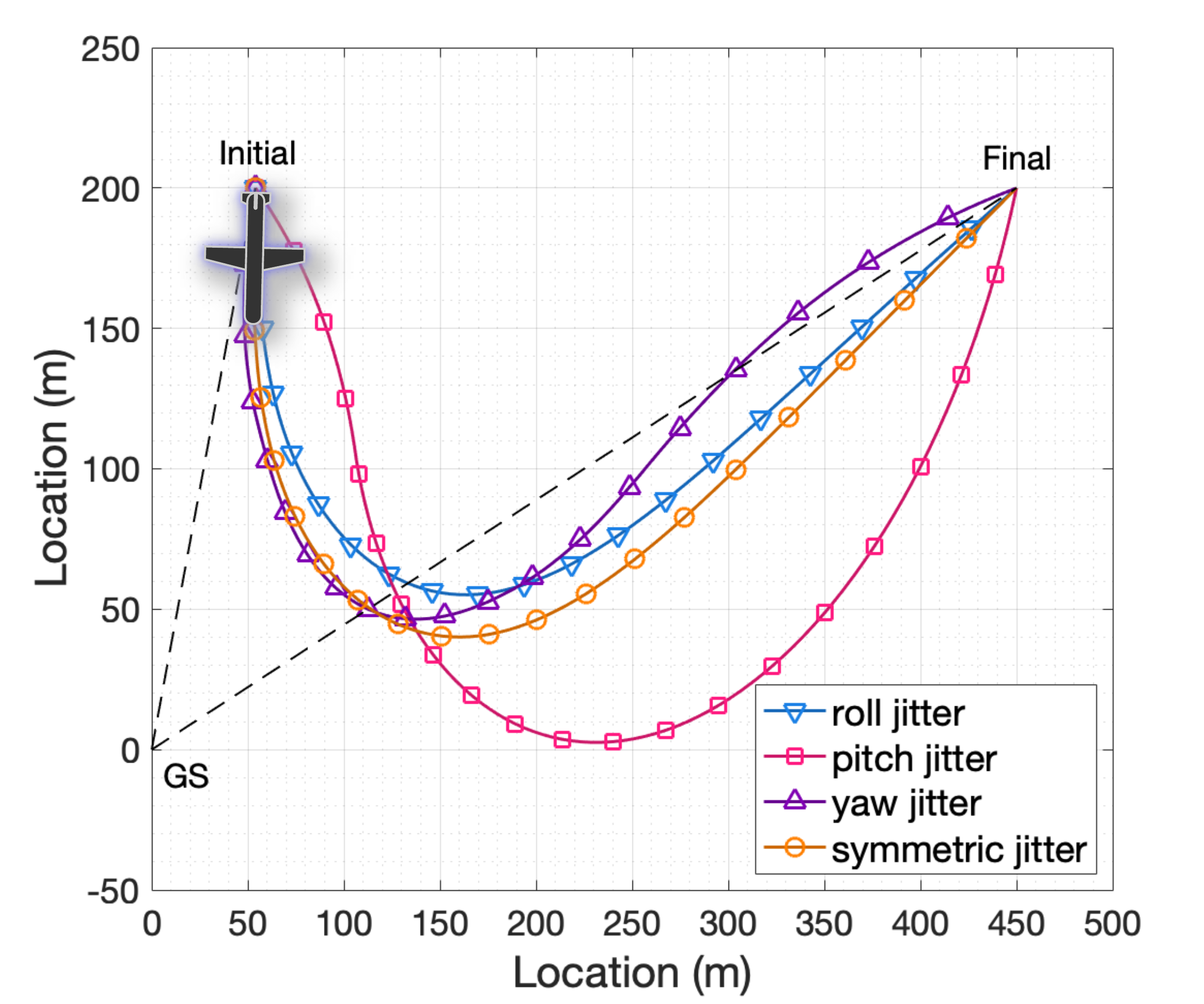}%
			%\caption{$1.5$ mrad}
			%\label{traj04}
		}
\subfloat[\label{line_traj_norm}]{%
        \includegraphics[width=0.62\columnwidth,keepaspectratio]
			{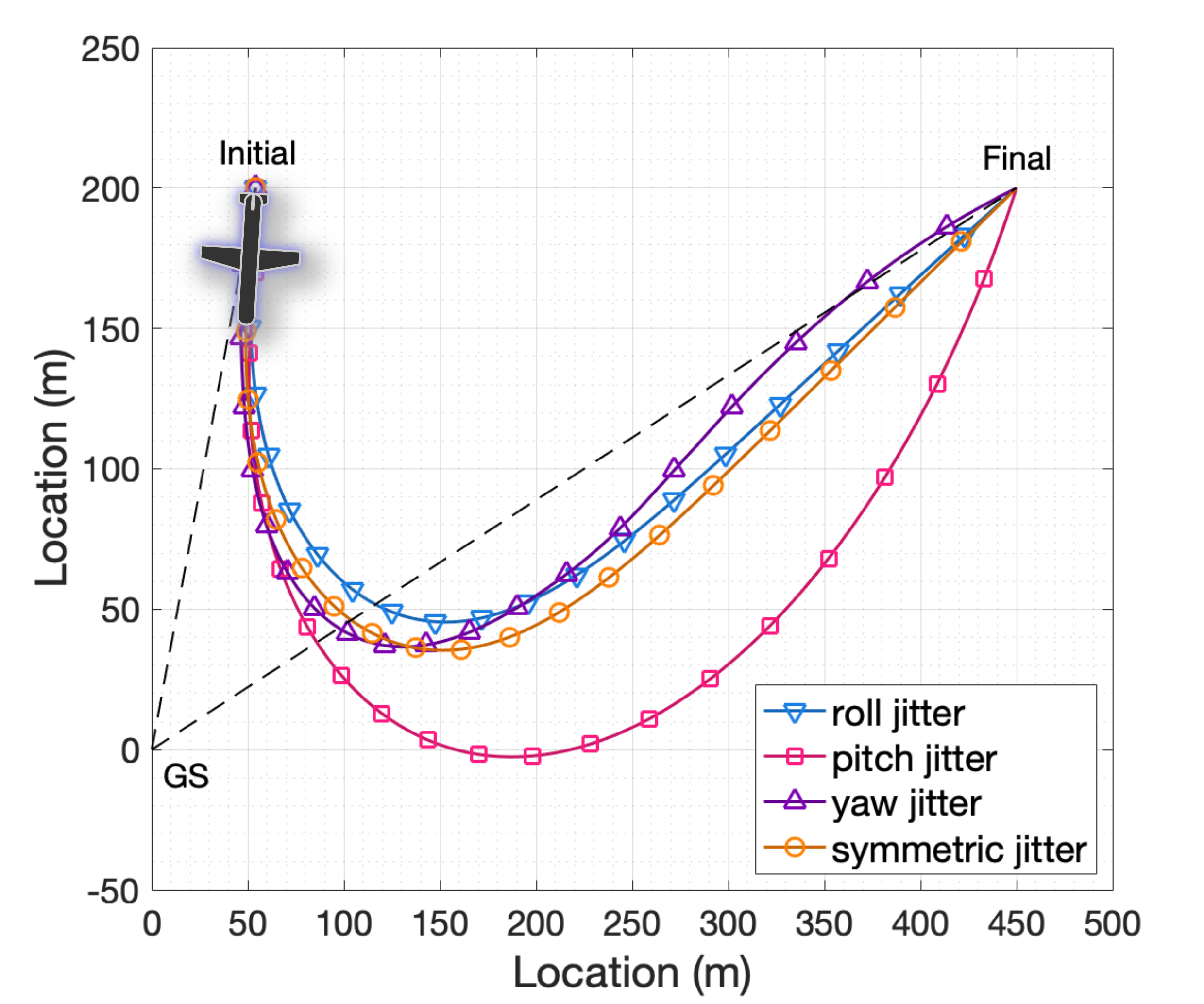}%
			%\caption{$2.0$ mrad}
			%\label{traj05}
		}
\subfloat[\label{line_traj_wide}]{%
        \includegraphics[width=0.62\columnwidth,keepaspectratio]
			{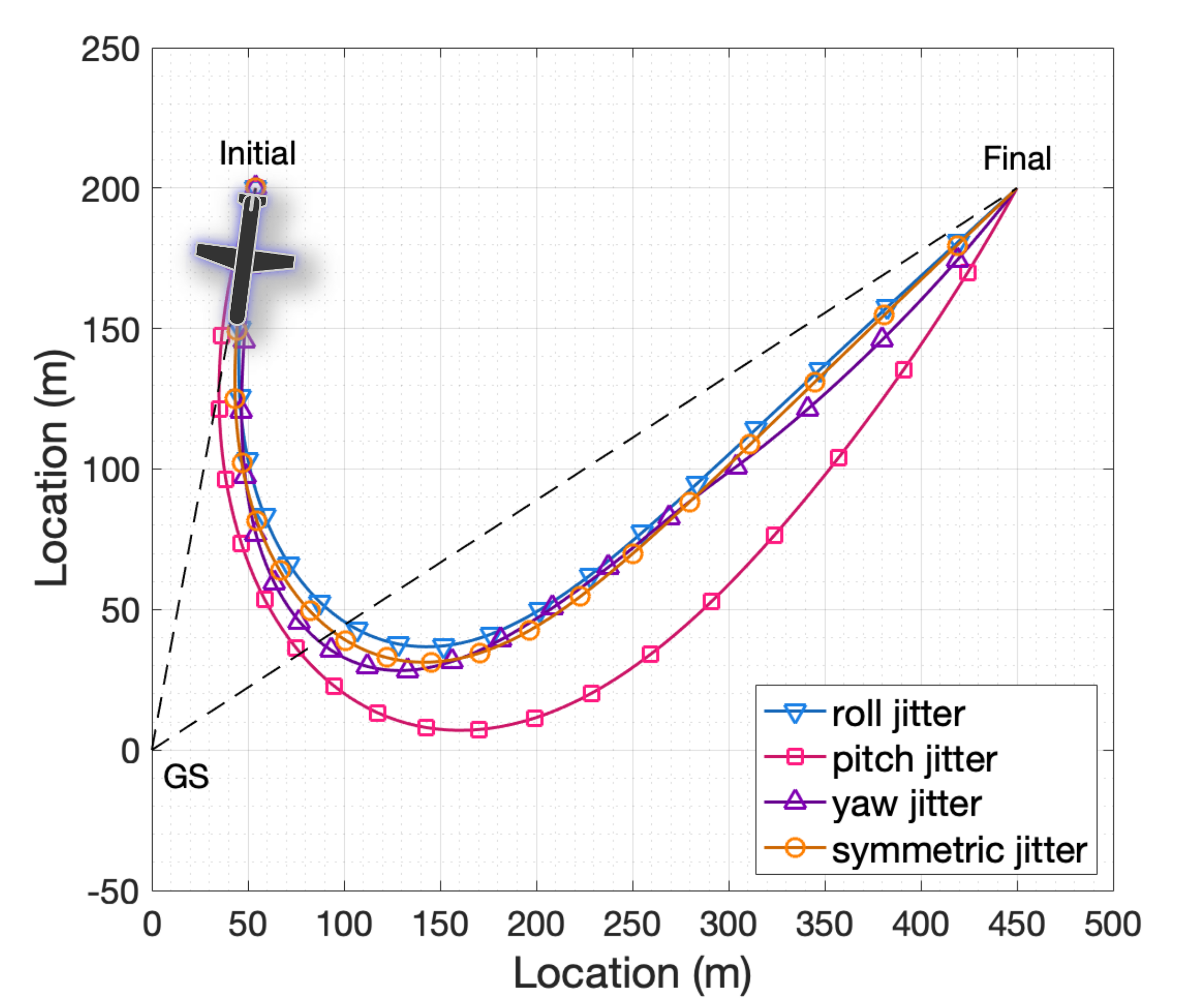}%
			%\caption{$2.5$ mrad}
			%\label{traj06}
		}
		\caption{The energy-efficient trajectories for the mission where the initial and final points are different (moving mission). We vary the parameter values by (a) $\sigma_\text{div} = 1.5$~mrad, (b) $\sigma_\text{div} = 2.0$~mrad, and (c) $\sigma_\text{div} = 2.5$~mrad to compare the impact of the pointing error on the trajectory.}
		\label{line_traj}
	%\end{center}
	\vspace{-12pt}
\end{figure*}

\begin{figure*}[t]
	%\begin{center}
	    \centering
\subfloat[\label{circ_traj_roll}]{%
       \includegraphics[width=0.62\columnwidth,keepaspectratio]
			{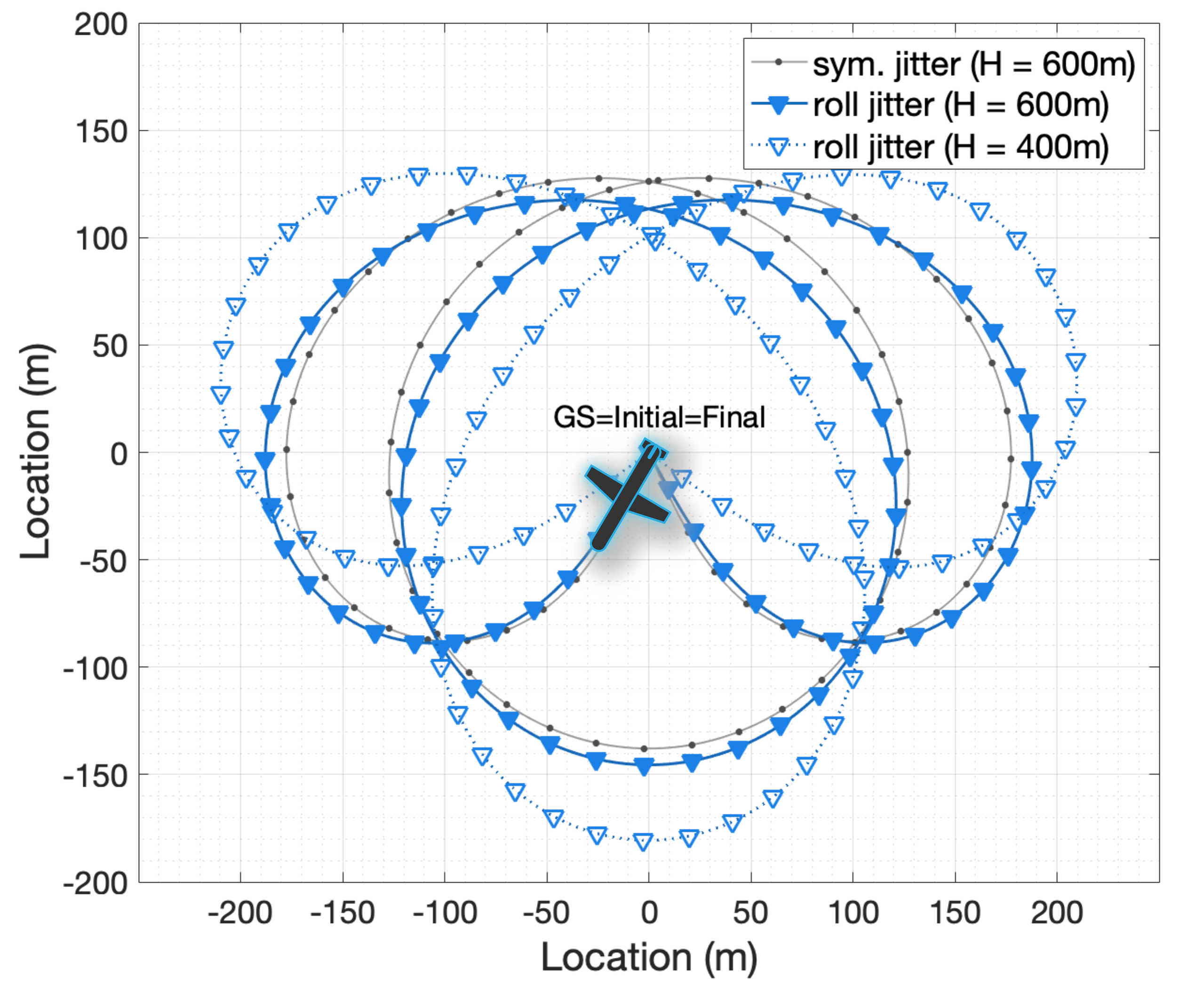}%
			%\caption{$1.5$ mrad}
			%\label{traj04}
		}
\subfloat[\label{circ_traj_pitch}]{%
        \includegraphics[width=0.62\columnwidth,keepaspectratio]
			{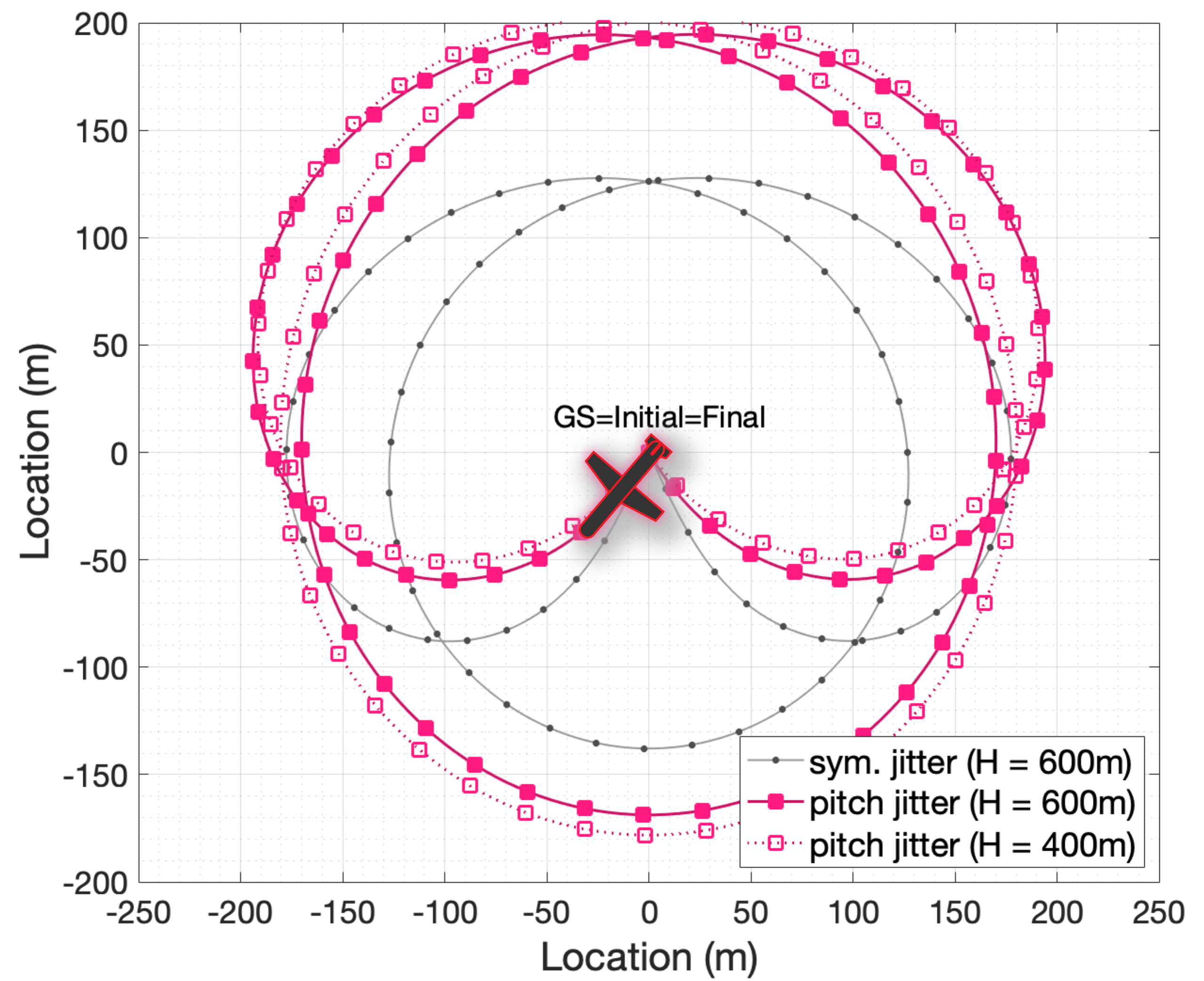}%
			%\caption{$2.0$ mrad}
			%\label{traj05}
		}
\subfloat[\label{circ_traj_yaw}]{%
        \includegraphics[width=0.62\columnwidth,keepaspectratio]
			{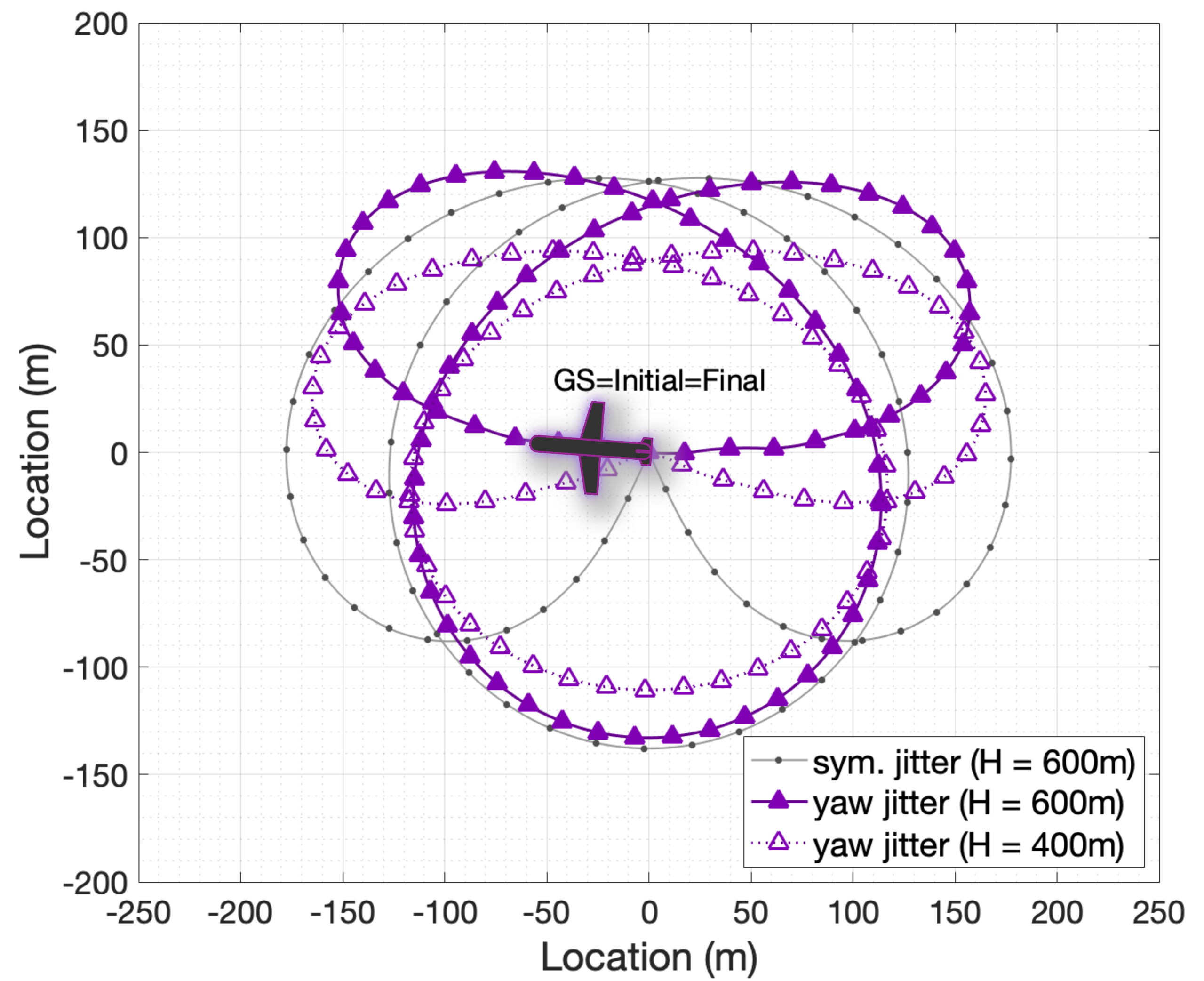}%
			%\caption{$2.5$ mrad}
			%\label{traj06}
		}
		\caption{The energy-efficient trajectories for the mission where the initial and final points are the same (hovering mission). The parameter is set to $\sigma_\text{div} = 1.5$~mrad, and each of the subfigures varies by the dominant jittering direction.}
		\label{circ_traj}
	%\end{center}
	\vspace{-12pt}
\end{figure*}

The constraints~(\ref{vnc1}), (\ref{nc1}), (\ref{nc6}), and (\ref{nc7}) can be approximated by the following inequalities~\cite{202003TVT}. The non-convex constraint (\ref{vnc1}) is restricted to
\begin{equation}
\label{ca9}
\begin{aligned}
\norm{\bold{v}[k]}\leq{{v}_\text{max}},\quad1\leq{k}\leq{N-1},
\end{aligned}
\end{equation}
\begin{equation}
\label{ca0}
\begin{aligned}
2{\bold{v}^{(p)}[k]}^T\bold{v}[k]-\norm{\bold{v}^{(p)}[k]}^2\geq{v_\text{min}^2},\quad1\leq{k}\leq{N-1},
\end{aligned}
\end{equation}
which is tight for $\bold{v}[k]=\bold{v}^{(p)}[k]$. The non-convex constraint (\ref{nc1}) is restricted to
\begin{equation}
\label{ca1}
\begin{aligned}
2{\bold{s}^{(p)}[k]}^T\bold{s}[k]-\norm{\bold{s}^{(p)}[k]}^2\geq{S_k^2},\quad\forall{k},
\end{aligned}
\end{equation}
which is tight for $\bold{s}[k]=\bold{s}^{(p)}[k]$. The non-convex constraint (\ref{nc6}) is restricted to
\begin{equation}
\label{ca4}
\begin{aligned}
2{\bold{v}^{(p)}[k]}^T\bold{v}[k]-\norm{\bold{v}^{(p)}[k]}^2\geq{R_k^2},\quad1\leq{k}\leq{N-1},
\end{aligned}
\end{equation}
which is tight for $\bold{v}[k]=\bold{v}^{(p)}[k]$. The non-convex constraint (\ref{nc7}) is restricted to
\begin{equation}
\label{ca5}
\begin{aligned}
\norm{Q_k-R_k,Q_k^{(p)}+R_k^{(p)},L_k-1,2,\frac{2\bold{a}[k]}{\textsl{g}}}&\leq{L_k+1},\\{\quad}1\leq{k}&\leq{N-1},
\end{aligned}
\end{equation}
where $L_k={(Q_k^{(p)}+R_k^{(p)})(Q_k+R_k)}/{2}$. The approximated constraint~(\ref{ca5}) is tight for $Q_k=Q_k^{(p)}$ and $R_k=R_k^{(p)}$. The non-convex constraint~(\ref{nc2}) is approximated as the following theorem.
\begin{theorem}
\label{theo3}
The non-convex constraint (\ref{nc2}) can be approximated into the convex constraint as
\begin{equation}
\label{ca6}
\begin{aligned}
S_k^{(p)}&U_k+U_k^{(p)}S_k-S_k^{(p)}U_k^{(p)}\geq\left({\hat{\bold{u}}^{(p)}[k]}^T\bold{D}\hat{\bold{u}}^{(p)}[k]\right)^\frac{1}{2}\\
&+\left({\hat{\bold{u}}^{(p)}[k]}^T\bold{D}\hat{\bold{u}}^{(p)}[k]\right)^{-\frac{1}{2}}\hat{\bold{u}}^{(p)}[k]\bold{D}\Delta\hat{\bold{u}}[k],\quad\forall{k},
\end{aligned}
\end{equation}
where $\bold{D}=\bold{\text{diag}}(\sigma_\beta^2+\sigma_\gamma^2,\sigma_\gamma^2+\sigma_\alpha^2,\sigma_\alpha^2+\sigma_\beta^2)$. The symbols $\hat{\bold{u}}^{(p)}[k]$ and $\Delta\hat{\bold{u}}[k]$ are expressed as~(\ref{hatup}) and (\ref{deltau}), respectively, where
\begin{figure*}[b!]
\vspace{-5pt}
\begin{center}
\line(1,0){515}
\end{center}
\vspace{3pt}
\begin{equation}
\label{hatup}
\begin{aligned}
\hat{\bold{u}}^{(p)}[k]=
\begin{bmatrix}
s_x^{(p)}\cos\theta^{(p)}+s_y^{(p)}\sin\theta^{(p)}\\[2pt]
-s_x^{(p)}\cos\phi^{(p)}\sin\theta^{(p)}+s_y^{(p)}\cos\phi^{(p)}\cos\theta^{(p)}+s_z^{(p)}\sin\phi^{(p)}\\[2pt]
-s_x^{(p)}\sin\phi^{(p)}\sin\theta^{(p)}+s_y^{(p)}\sin\phi^{(p)}\cos\theta^{(p)}+s_z^{(p)}\cos\phi^{(p)}
\end{bmatrix}
\end{aligned}
\end{equation}
\begin{equation}
\label{deltau}
\begin{aligned}
\Delta\hat{\bold{u}}[k]=
\begin{bmatrix}
s_x^{(p)}\Delta\cos\theta+\cos\theta^{(p)}\Delta{s_x}+s_y^{(p)}\Delta\sin\theta+\sin\theta^{(p)}\Delta{s_y}\\[3pt]
-s_x^{(p)}\cos\phi^{(p)}\Delta\sin\theta-s_x^{(p)}\sin\theta^{(p)}\Delta\cos\phi-\cos\phi^{(p)}\sin\theta^{(p)}\Delta{s_x}+s_y^{(p)}\cos\phi^{(p)}\Delta\cos\theta\\
+s_y^{(p)}\cos\theta^{(p)}\Delta\cos\phi+\cos\phi^{(p)}\cos\theta^{(p)}\Delta{s_y}+s_z^{(p)}\Delta\sin\phi+\sin\phi^{(p)}\Delta{s_z}\\[3pt]
-s_x^{(p)}\sin\phi^{(p)}\Delta\sin\theta-s_x^{(p)}\sin\theta^{(p)}\Delta\sin\phi-\sin\phi^{(p)}\sin\theta^{(p)}\Delta{s_x}+s_y^{(p)}\sin\phi^{(p)}\Delta\cos\theta\\
+s_y^{(p)}\cos\theta^{(p)}\Delta\sin\phi+\sin\phi^{(p)}\cos\theta^{(p)}\Delta{s_y}+s_z^{(p)}\Delta\cos\phi+\cos\phi^{(p)}\Delta{s_z}
\end{bmatrix}
\end{aligned}
\end{equation}
\begin{equation}
\label{cosphi}
\begin{aligned}
\Delta\sin\phi=(a_0^2+1)^{-3/2}\norm{\bold{v}^{(p)}[k]}^{-3}\{(v_y^{(p)2}a_y^{(p)}-v_x^{(p)}v_y^{(p)}a_x^{(p)})\Delta{v_x}+(v_x^{(p)2}a_x^{(p)}-v_x^{(p)}v_y^{(p)}a_y^{(p)})\Delta{v_x}\\[-1pt]
+v_x^{(p)}(v_x^{(p)2}+v_y^{(p)2})\Delta{a_y}-v_y^{(p)}(v_x^{(p)2}+v_y^{(p)2})\Delta{a_x}\},\\
\Delta\cos\phi=a_0(a_0^2+1)^{-3/2}\norm{\bold{v}^{(p)}[k]}^{-3}\{(v_y^{(p)2}a_y^{(p)}-v_x^{(p)}v_y^{(p)}a_x^{(p)})\Delta{v_x}+(v_x^{(p)2}a_x^{(p)}-v_x^{(p)}v_y^{(p)}a_y^{(p)})\Delta{v_x}\\[-1pt]
+v_x^{(p)}(v_x^{(p)2}+v_y^{(p)2})\Delta{a_y}-v_y^{(p)}(v_x^{(p)2}+v_y^{(p)2})\Delta{a_x}\}
\end{aligned}
\end{equation}\noindent
\end{figure*}\noindent
\begin{equation}
\label{sintheta}
\begin{aligned}
\cos\theta^{(p)}=\frac{v_x^{(p)}}{\norm{\bold{v}^{(p)}[k]}},\,\,\sin\theta^{(p)}=\frac{v_y^{(p)}}{\norm{\bold{v}^{(p)}[k]}}
\end{aligned}
\end{equation}
\begin{equation}
\label{costheta}
\begin{aligned}
\Delta\sin\theta=\norm{\bold{v}^{(p)}[k]}^{-3}(v_y^{(p)2}\Delta{v_x}-v_x^{(p)}v_y^{(p)}\Delta{v_y}),\\
\Delta\cos\theta=\norm{\bold{v}^{(p)}[k]}^{-3}(v_x^{(p)2}\Delta{v_y}-v_x^{(p)}v_y^{(p)}\Delta{v_x}),
\end{aligned}
\end{equation}
\begin{equation}
\label{sinphi}
\begin{aligned}
\sin\phi^{(p)}=\frac{a_0}{\sqrt{a_0^2+1}},\,\,\cos\phi^{(p)}=\frac{1}{\sqrt{a_0^2+1}},
\end{aligned}
\end{equation}
and (\ref{cosphi}). The symbol $a_0$ is expressed as follows:
\begin{equation}
\label{azero}
\begin{aligned}
a_0=\frac{v_y^{(p)}a_x^{(p)}-v_x^{(p)}a_y^{(p)}}{\norm{\bold{v}[k]}\textsl{g}}.
\end{aligned}
\end{equation}
In (\ref{deltau}), $\Delta\hat{\bold{u}}[k]$ is approximated into the linear combination of ${\Delta}s_x,{\Delta}s_y,{\Delta}v_x,{\Delta}v_y,{\Delta}a_x,{\Delta}a_y$, where ${\Delta}s_x=s_x-s_x^{(p)}$, ${\Delta}s_y=s_y-s_y^{(p)}$, ${\Delta}v_x=v_x-v_x^{(p)}$, ${\Delta}v_y=v_y-v_y^{(p)}$, ${\Delta}a_x=a_x-a_x^{(p)}$, ${\Delta}a_y=a_y-a_y^{(p)}$. The approximated constraint~(\ref{ca6}) is also tight for the variable values of the $p$-th iteration.
\end{theorem}

\proof
See Appendix~\ref{appen3}.
\endproof

\begin{lemma}
\label{theo4}
The constraint~(\ref{nc3}) is convex under (\ref{elev2}).
\end{lemma}

\proof
See Appendix~\ref{appen4}.
\endproof

In addition, although (\ref{nc3}) is a non-convex constraint, it is not necessary to transform it into a convex constraint according to \emph{Lemma}~\ref{theo4}.

	\subsection{Dinkelbach Method}
	\label{dinkelbachsection}

In Section~\ref{scasection}, the optimization problem with the concave-convex fractional objective function is formulated as (P2), and the non-convex constraints are approximated to~(\ref{ca9})-(\ref{ca6}). We formulate the final problem to iteratively solve (P2) by the Dinkelbach method~\cite{1967MS,2010FTML,202102TWC}. The total capacity and power consumption are denoted as $C_\text{tot}$ and $P_\text{tot}$, respectively, and can be represented as
\begin{equation}
\label{totalthings}
\begin{aligned}
{C}_\text{tot}&=\sum_{k=1}^{N}\nabla_{\text{L},k}^{(p)}(c_3-c_4\norm{\bold{s}[k]}-c_5\check{U}_k^2-c_6\check{V}_k)+\delta_{\text{L},k}^{(p)},\\
{P}_\text{tot}&=\sum_{k=1}^{N-1}\check{P}_k+NP_\text{T}+{E_\text{cost}}/{\delta},
\end{aligned}
\end{equation}
derived from~(\ref{obj2}). Then, we define (P3) as follows:
\begin{subequations}
\label{problemdink}
\begin{align}
\text{(P3)}:\notag\\
\begin{split}
\label{obj3}
\min_{\bold{s}[k]}\quad&{F}=-{C}_\text{tot}+\lambda{P}_\text{tot}
\end{split}
\\
\begin{split}
\label{con3}
\textrm{s.t.}\quad&(\ref{convexc}), (\ref{convexd}), (\ref{nc2})-(\ref{nc5}), (\ref{elev2}), (\ref{ca9})-(\ref{ca6})
\end{split}
\end{align}
\end{subequations}

Algorithm~\ref{alg1} provides the suboptimal trajectory of the fixed-wing UAV that improves the energy efficiency. First, the original and auxiliary variables are initialized to formulate the approximated problem and conduct the SCA method. The feasible set of (P2) can be approximated into convex. However, (\ref{obj2}) is still non-convex. Therefore, we apply the Dinkelbach method to iteratively solve (P3) and obtain the solution for the $p$-th iteration of the SCA method.

\subsection{Convergence and Complexity Analysis}

The SCA method guarantees convergence only when the feasible set of the approximated problem is both convex and included in the feasible set of the original problem~\cite{2016OE}. However,~(\ref{ca6}) is convex but not included in the solution set of~(\ref{nc2}). Nevertheless, the right-hand side (RHS) of~(\ref{nc2}) is a smooth nonlinear function (compared to other constraints) with respect to $\bold{s}[k]$, $\bold{v}[k]$, and $\bold{a}[k]$, and it is non-negative, meaning that the equality of~(\ref{nc2}) always holds since $\check{U}_k$ must be minimized. As a result, although there is no general convex restriction for~(\ref{nc2}), the proposed approximation into the convex constraint~(\ref{ca6}) closely approximates the value of $\check{U}_k^2$ in~(\ref{obj2}) and does not compromise the physical feasibility.
%Furthermore, the RHS of~(\ref{ca6}) can be considered as a positive for two reasons. First, non-negative property is hold for a very wide range of $\Delta\hat{\bold{u}}[k]$, as $\norm{\Delta\hat{\bold{u}}[k]}\leq\norm{\hat{\bold{u}}^{(p)}[k]}$. Second, if the optimal solution during the iteration is found in $\norm{\Delta\hat{\bold{u}}[k]}>\norm{\hat{\bold{u}}^{(p)}[k]}$, the constraint can be ignored since the best choice is trivially a $\check{U}_k=0$.
In summary, although Algorithm~\ref{alg1} may not result in a converged solution, it only results in a slight oscillation of the objective value without resulting in infeasible solutions. Moreover, since all constraints in (P3), including~(\ref{ca6}), are tight for the variable values of the $p$-th iteration, the algorithm converges in most cases.

The computational complexity of our proposed algorithm can be evaluated based on the complexity of solving the convex optimization problem (P3). The number of iterations required can vary depending on factors such as the type of objective and constraint functions, threshold values, and learning rates. For a convex optimization problem, the interior-point method provides a generalized computational complexity of $\mathcal{O}(\max\{n_\text{v}^3, n_\text{v}^2 n_\text{c}, F\})$, where $n_v$ represents the number of variables, $n_c$ represents the number of constraints, and $F$ represents the cost of evaluating the first and second derivatives of the objective function and constraints~\cite{202101TWC,boyd}. Despite the complicated formulation of the constraints, particularly for~(\ref{ca6}), $F$ follows $\mathcal{O}(n_\text{v})$, suggesting it can generally be ignored. The optimization variables are the $x$ and $y$ components of the vectors $\bold{s}[k]$, so $n_\text{v} = 2N$. The number of constraints related to $(\ref{convexc})$, $(\ref{convexd})$, $(\ref{nc2})-(\ref{nc5})$, $(\ref{elev2})$, and $(\ref{ca9})-(\ref{ca6})$ are $N-1$, $2N-1$, $N$, $N$, $4$, $N-1$, $N$, $N-1$, $N-1$, $N$, $N-1$, $N-1$, and $N$ respectively. Consequently, the total number of constraints is $n_\text{c} = 13N - 3$. Therefore, each iteration of our optimization process has a polynomial computational complexity of $\mathcal{O}(N^3)$.

	\section{Numerical Results}
	\label{numressection}

\begin{figure}[t]
	\begin{center}
		{\includegraphics[width=0.9\columnwidth,keepaspectratio]
			{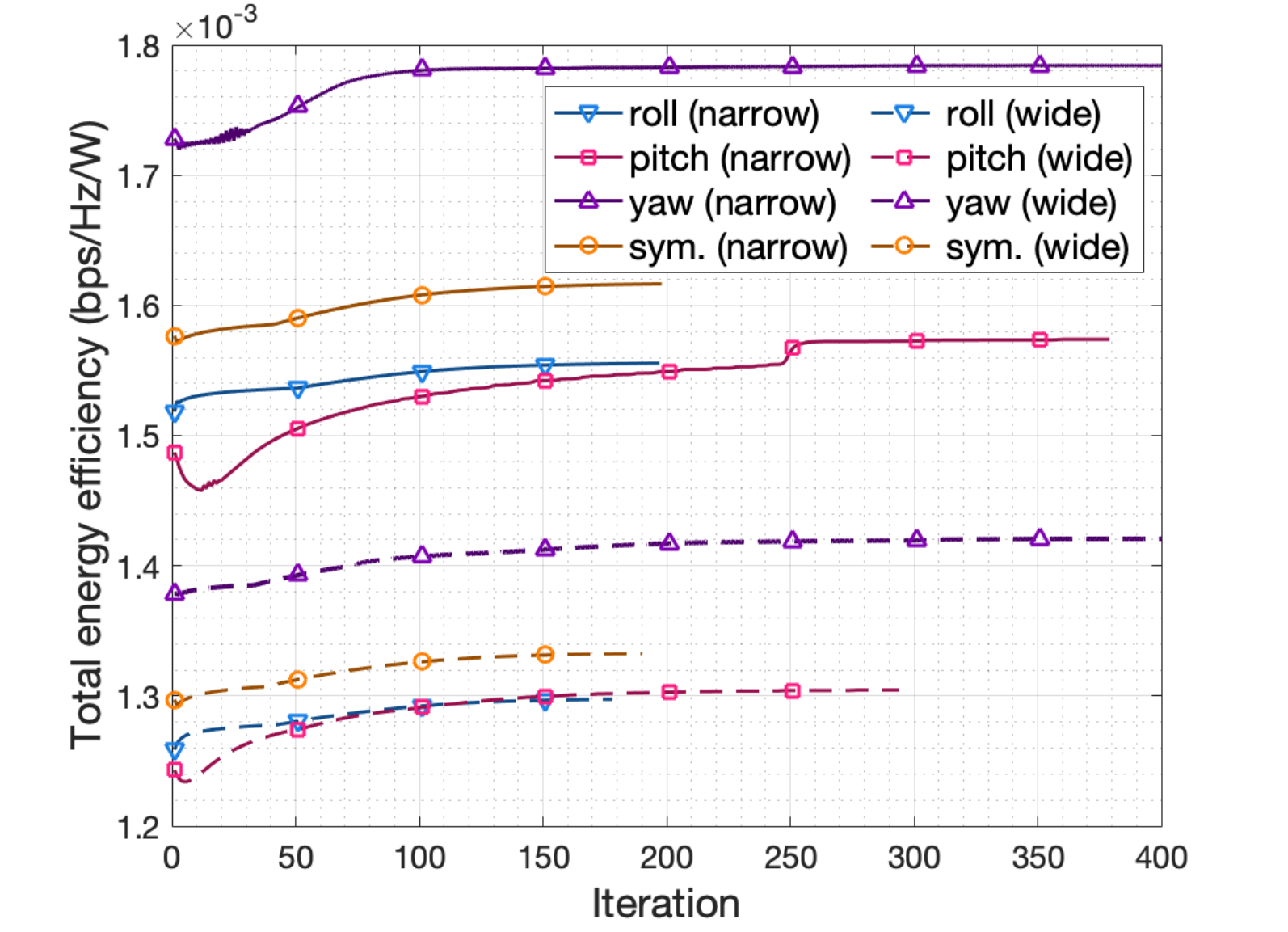}%
			\caption{Convergence of the SCA method (moving mission).}
			\label{line_conv}
		}
	\end{center}
	\vspace{-12pt}
\end{figure}

\begin{figure}[t]
	\begin{center}
		{\includegraphics[width=0.9\columnwidth,keepaspectratio]
			{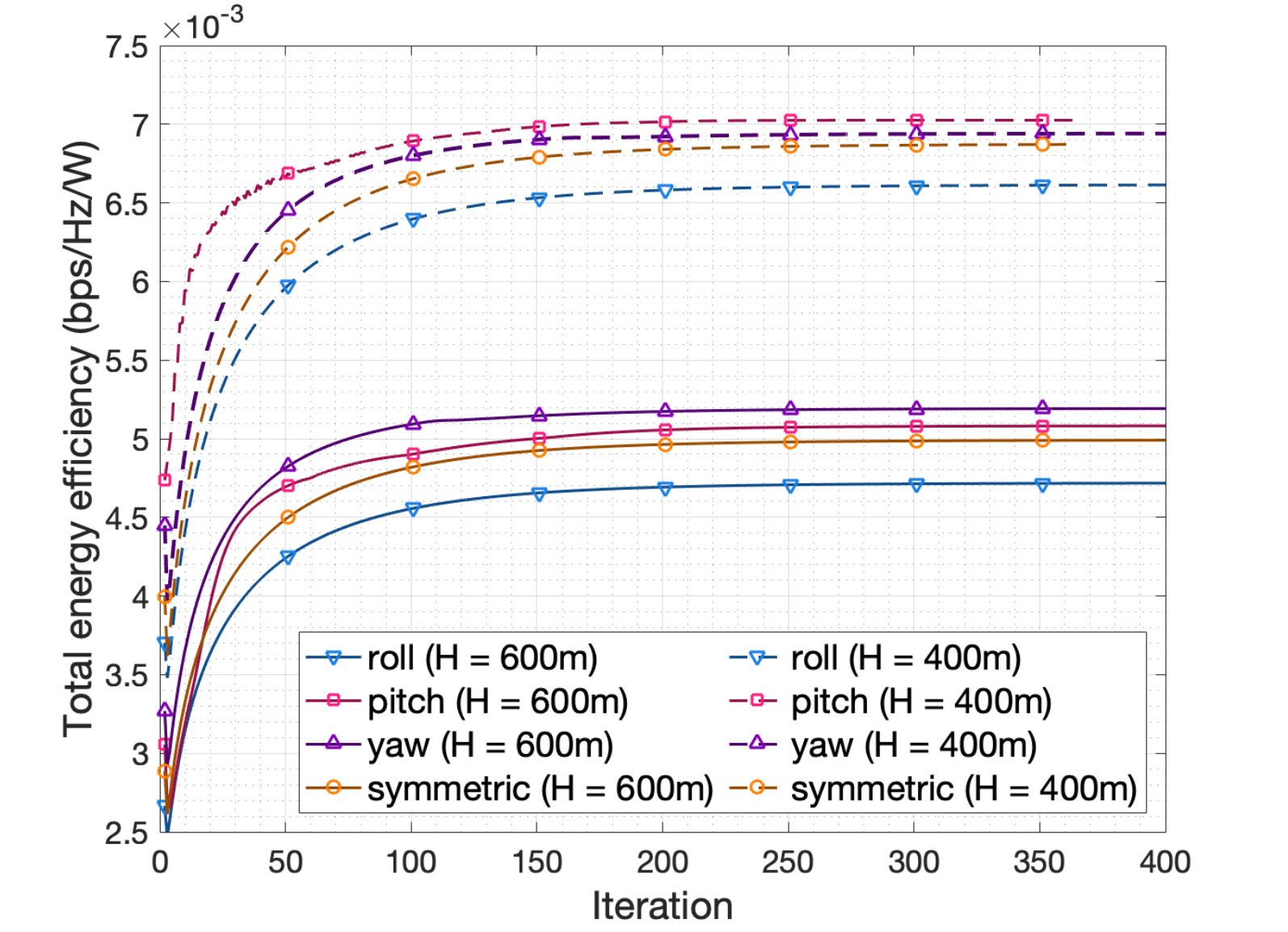}%
			\caption{Convergence of the SCA method (hovering mission).}
			\label{circ_conv}
		}
	\end{center}
	\vspace{-12pt}
\end{figure}

\begin{comment}
\begin{figure}[t]
	\begin{center}
		{\includegraphics[width=0.9\columnwidth,keepaspectratio]
			{Fig/traj08_03.pdf}%
			\caption{Spectral efficiency of the trajectories for the moving mission over time.}
			\label{line_spec}
		}
	\end{center}
	\vspace{-12pt}
\end{figure}
\end{comment}

\begin{figure}[t]
	\begin{center}
		{\includegraphics[width=0.9\columnwidth,keepaspectratio]
			{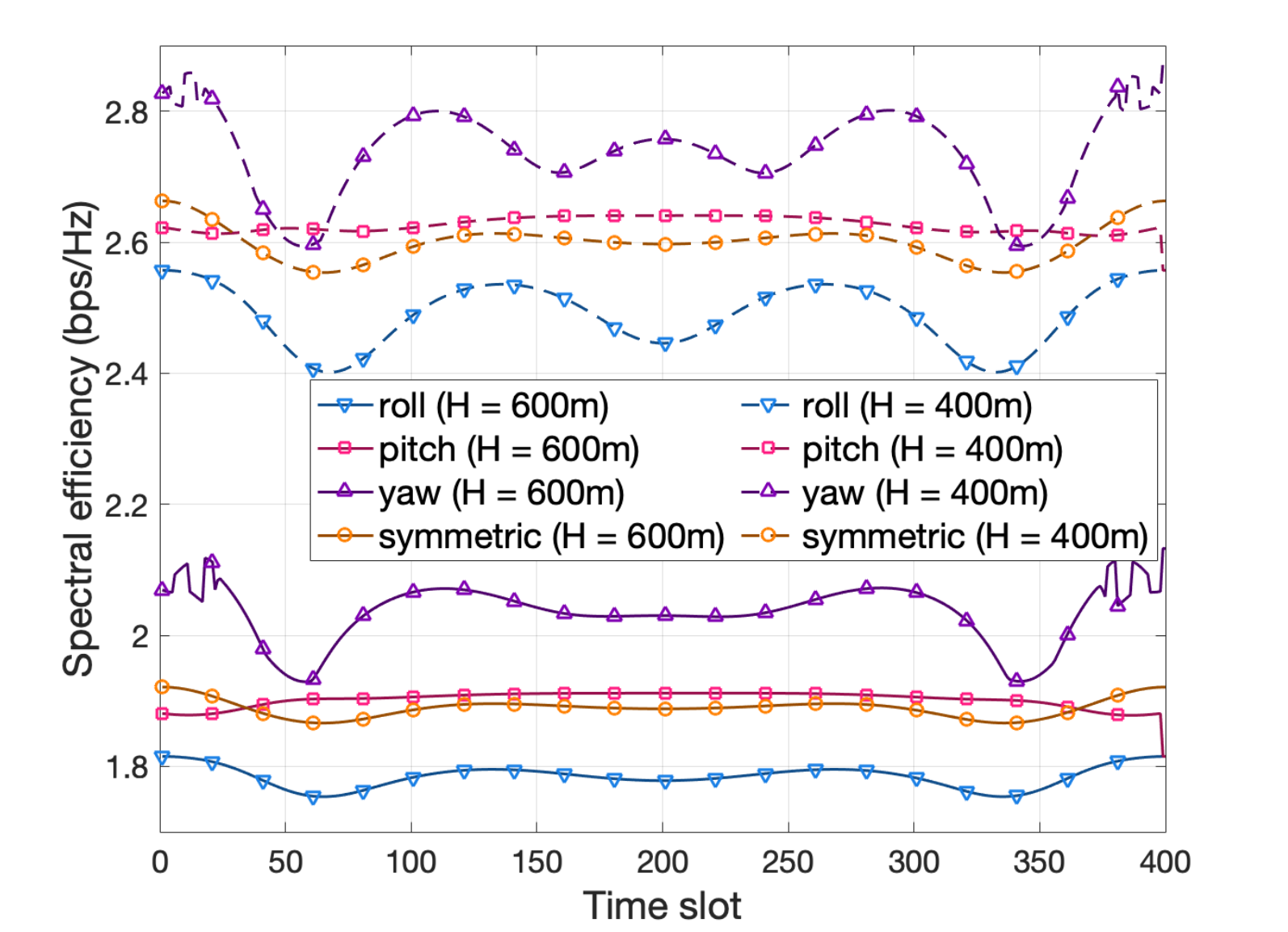}%
			\caption{Spectral efficiency of the trajectories over time (hovering mission).}
			\label{circ_spec}
		}
	\end{center}
	\vspace{-12pt}
\end{figure}

\begin{figure}[t]
	\begin{center}
		{\includegraphics[width=0.9\columnwidth,keepaspectratio]
			{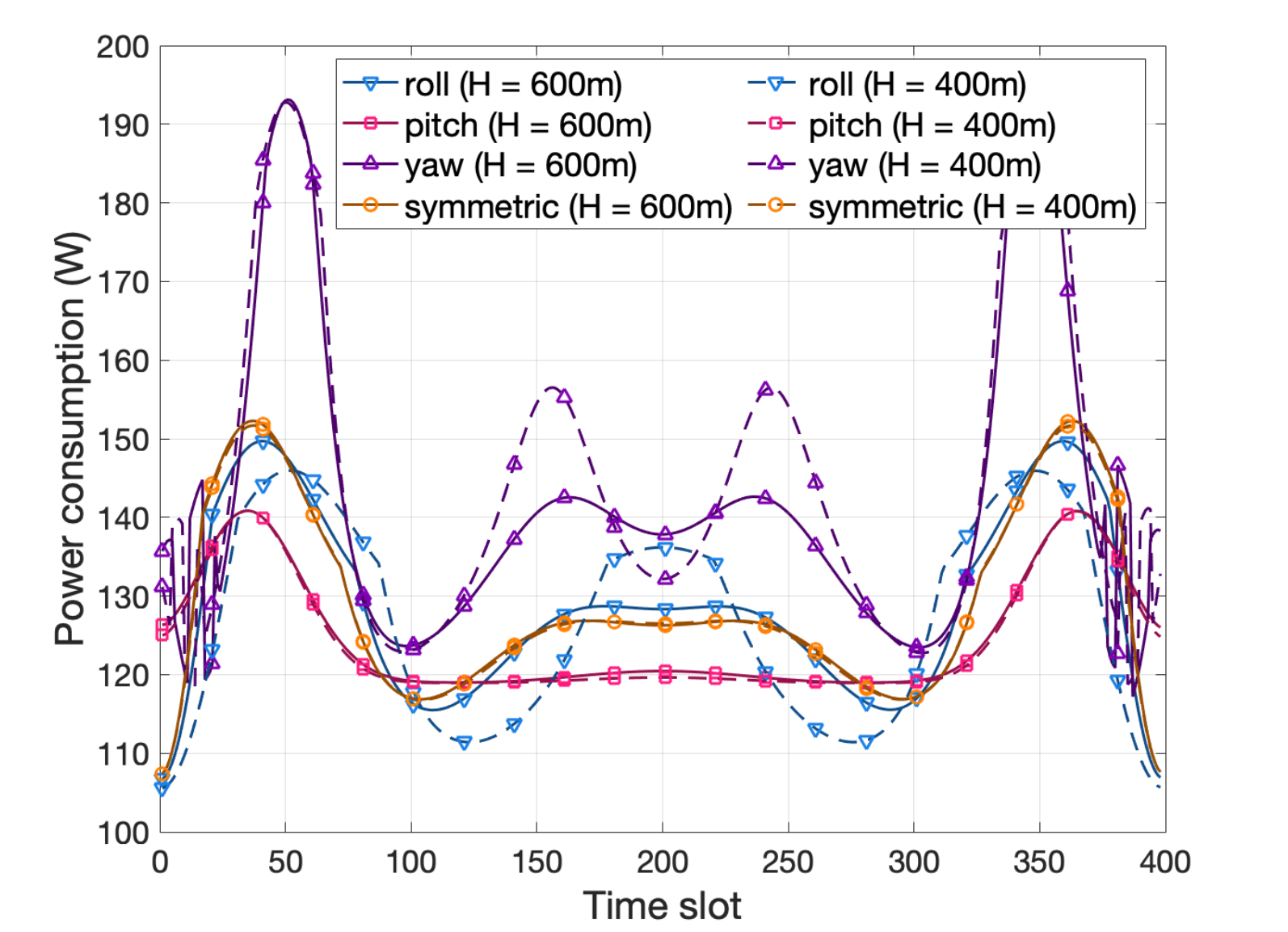}%
			\caption{Power consumption of the trajectories over time (hovering mission).}
			\label{circ_power}
		}
	\end{center}
	\vspace{-12pt}
\end{figure}

In this section, we present numerical results that validate the significance of the proposed pointing error model and the algorithm for designing an energy-efficient trajectory. The simulation parameters are specified in Table~\ref{tbl2}. The altitude of the UAV is set to $600$~m and $400$~m, and the transmit power is $10$~mW with $P_\text{T}/\sigma=30$~dB~\cite{202111JSAC,202003TWC}. The flight time is $20$~s and $80$~s for two different flight missions, and is discretized into time slots with a $0.2$~s interval. The minimum and maximum speeds of the UAV are $3$ and $100$~$\mathrm{m/s}$, respectively, and the maximum acceleration is $5$~$\mathrm{m/s^2}$. The acceleration constraint also limits the bank angle during flight to $|{\phi}|\leq{27.03}\cdots^\circ$~\cite{201706TWC}. To account for the fixed cost of the flight mission, we introduce the parameter $E_\text{cost}$. Increasing $E_\text{cost}$ to a larger value induces more variations in the trajectory. We set the value to $10^5$J and $4\times10^5$J for two different missions, respectively. We assume that the GS is located at $[0, 0, 0]$~m in the GS-centered coordinate system.

%%%%%%%%%%%%%%%%%%%%%%%%%%%%%%%%%%%% revised

%%%% Fig. 4

\subsection{Comparison Between Different Jitter Characteristics}

We begin by comparing the energy-efficient trajectory obtained from Algorithm~\ref{alg1} for different types of UAV jitter. Fig.~\ref{line_traj} and Fig.~\ref{circ_traj} illustrate the trajectory for dominant roll jitter, pitch jitter, yaw jitter, and symmetric jitter which are represented by $[\sigma_\alpha, \sigma_\beta, \sigma_\gamma]=[1,0.1,0.1],$ $[0.1,1,0.1],$ $[0.1,0.1,1],$ $[0.583,0.583,0.583]$~$\mathrm{mrad}$, respectively. In all cases, the total magnitude of jittering remains constant at $\sqrt{\text{Tr}(\Sigma)}=1.01$~$\mathrm{mrad}$. The covariance between the jittering directions is assumed to be zero ($\rho_{\alpha,\beta}=\rho_{\beta,\gamma}=\rho_{\gamma,\alpha}=0$). In Fig.~\ref{line_traj}, the trajectories of the moving mission are displayed, where the initial and final points of the mission are set at $[54, 200, 600]$~m and $[450, 200, 600]$~m, respectively. For the moving mission, we initialize the parameters with a uniform linear motion from the initial to the final point. Fig.~\ref{circ_traj}, the hovering mission, sets the initial and final points to $[0, 0, H]$~m, where $H=400,\,600$~m. For the hovering mission, we initialize the trajectory with a clockwise circular path centered at $[0,-60,H]$~m. The flight duration is $20$~s and $80$~s for the two different missions, respectively.

Fig.~\ref{line_traj_narr} exhibits largest difference between trajectories, since $\theta_p$ has a more significant impact on performance when $\sigma_\text{div}$ is small, as expressed in~(\ref{pointingloss}). For the symmetric jitter case, the UAV aims to approach the GS while reducing energy consumption through smooth flight. A high amount of pitch jitter significantly affects the energy-efficient trajectory. The UAV should keep the $y'$ axis parallel to the link direction to avoid the pitch jitter. Consequently, at the beginning of the flight, the UAV turns in a different direction compared to the other cases. The UAV also chooses to fly further later in the flight to utilize the bank angle caused by acceleration. When the roll jitter dominates, the UAV tends to align the $x'$ axis with the line connecting the GS and the final point. Finally, for the yaw jitter case, the UAV accelerates at the end of the flight to minimize the pointing error. Similar differences are also shown in Fig.~\ref{line_traj_norm} and Fig.~\ref{line_traj_wide}. However, as $\sigma_\text{div}$ increases, trajectories optimized for asymmetric jitter characteristics begin to converge towards those optimized for symmetric jitter characteristics. We also present the simulation results of the hovering trajectory of the UAVs in Fig.~\ref{circ_traj}. We compare the energy-efficient trajectory for different types of jittering features when the initial and final points are the same.

\begin{figure}[t]
	\begin{center}
		{\includegraphics[width=0.92\columnwidth,keepaspectratio]
			{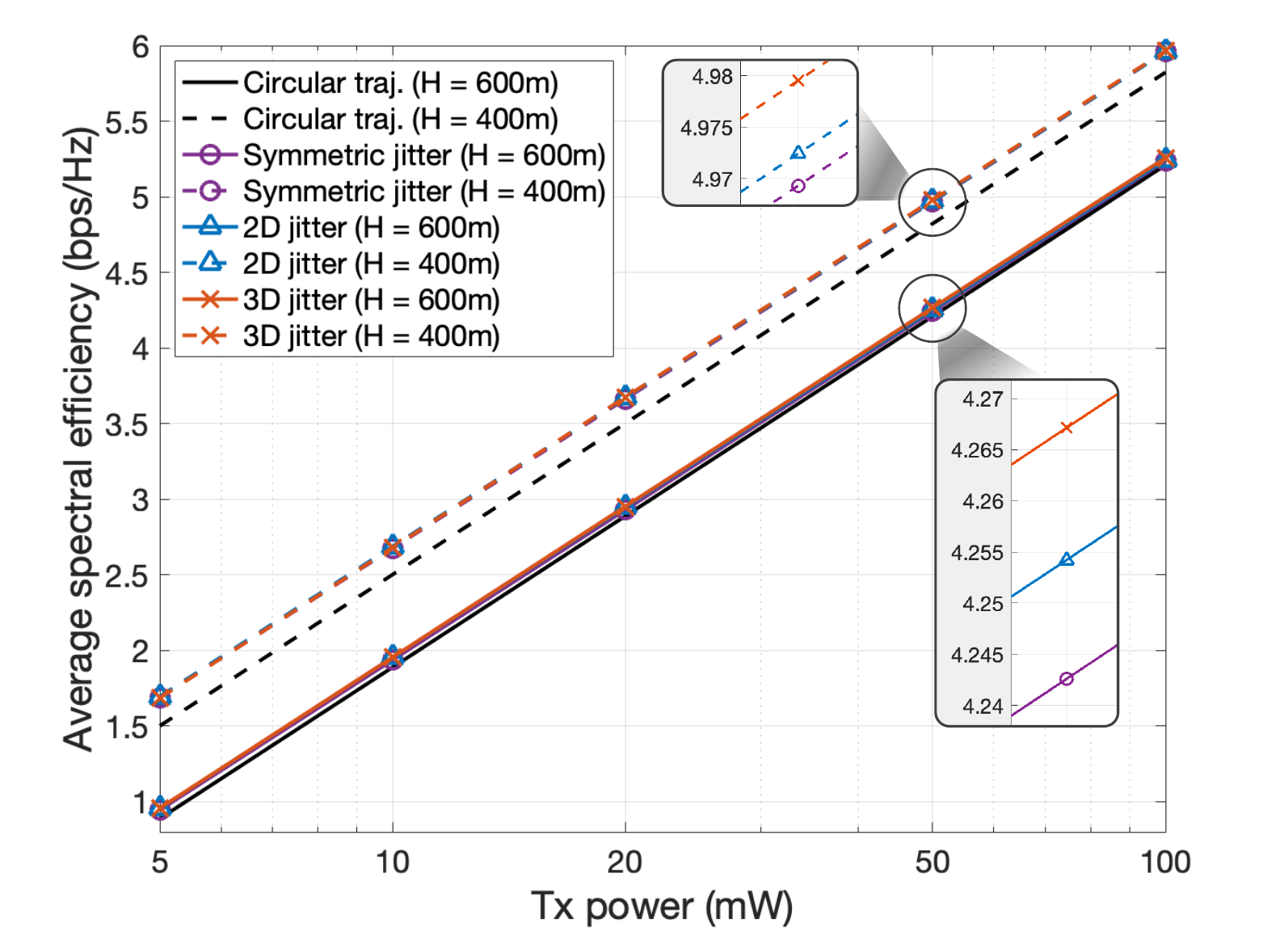}%
			\caption{Average spectral efficiency of trajectories derived using different jitter models.}
			\label{fig10}
		}
	\end{center}
	\vspace{-12pt}
\end{figure}

%%%% Fig. 6 and 7

In Fig.~\ref{line_conv} and Fig.~\ref{circ_conv}, we present the convergence of the SCA method for the two different missions: the moving mission and the hovering mission. For each iteration, we calculate the total energy efficiency, represented as $C_\text{tot}/P_\text{tot}$. Iteration stops when convergence is detected by the algorithm. Generally, the proposed algorithm converges effectively; however, slight oscillations are observed in some cases with dominant yaw jitter. This occurs because the feasible set of~(\ref{ca6}) is not included in the original feasible set. Nevertheless, as discussed in Section~\ref{dinkelbachsection}, the iterative solutions do not violate the physical feasibility defined in the system model, even during the oscillation.

%%%% Fig. 8, 9, and 10

%In Fig.~\ref{line_spec}, we show the value of $C_\mathbb{E}[k]$ during flight for the trajectories of the moving mission in Fig.~\ref{line_traj_narr} and Fig.~\ref{line_traj_norm}. The link distance is the most critical factor determining the spectral efficiency, so the trajectories tend to get closer to the GS in the middle of the flight. Thus, the spectral efficiency is generally highest when the UAV is closer to the origin. In contrast, 
The spectral efficiency presented in Fig.~\ref{circ_spec} over the trajectories of the hovering mission illustrates the strong impact of the posture of the UAV on the pointing error. Especially for the pitch jitter case, the corresponding trajectory depicted in Fig.~\ref{circ_traj_pitch} demonstrates that the $y'$ axis of the UAV is aligned with the ground station (GS) due to the acceleration during the large circular flight, except for the beginning and end of the flight. As a result, although the UAV is closest to the GS, the spectral efficiency is lower at both ends of the flight. Finally, in Fig.~\ref{circ_power}, we display the value of $P_\text{F}(\bold{v}[k],\bold{a}[k])$ for the moving mission. In our results, the UAV generally has a higher speed and acceleration for the pitch jitter case. However, the highest power consumption was found in the yaw jitter case, which is in $10$~s and $70$~s regions with high acceleration and relatively low speed.

\begin{figure}[t]
	\begin{center}
		{\includegraphics[width=0.92\columnwidth,keepaspectratio]
			{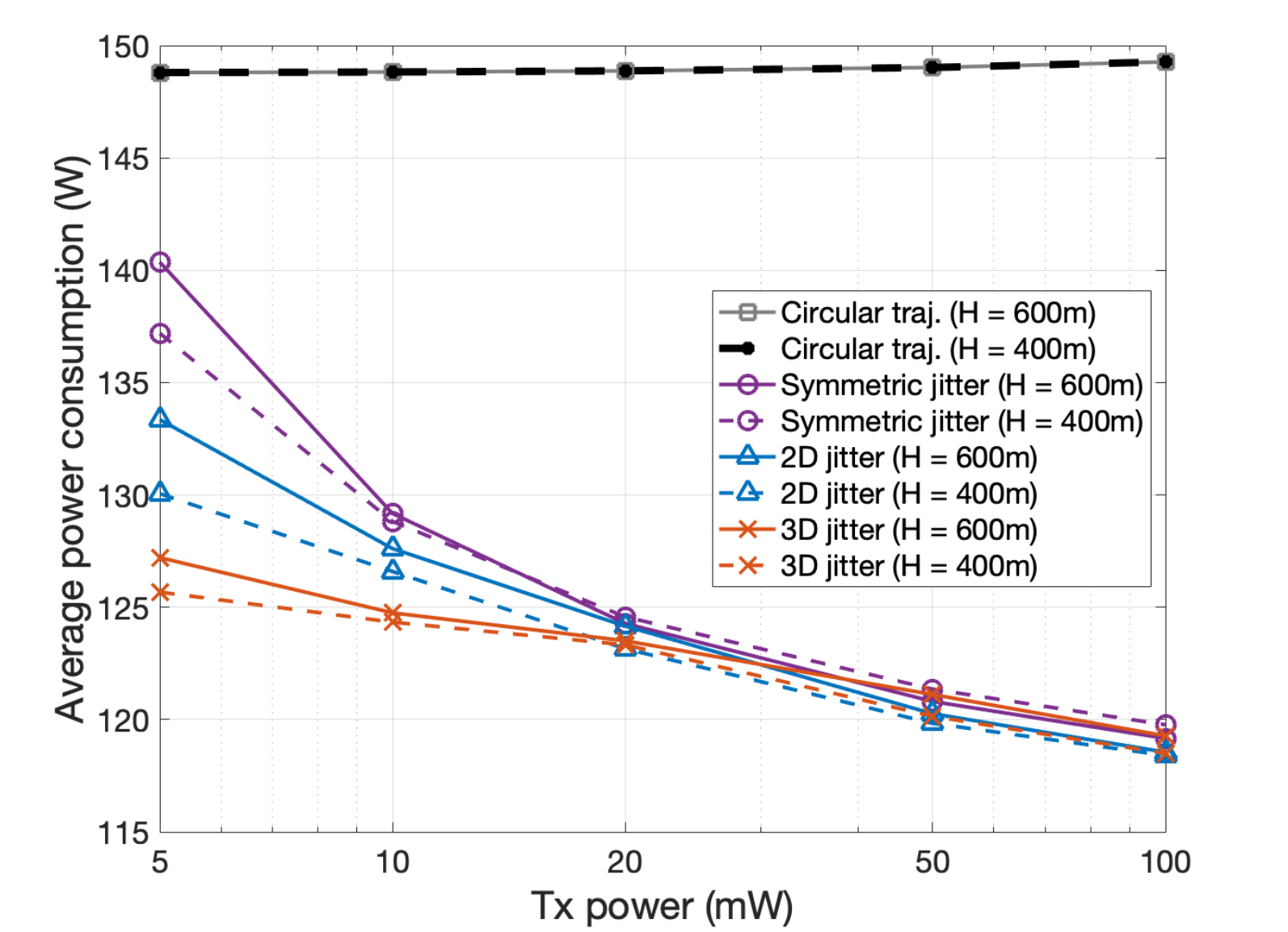}%
			\caption{Average power consumption of trajectories derived using different jitter models.}
			\label{fig11}
		}
	\end{center}
	\vspace{-12pt}
\end{figure}

\begin{figure*}[t]
	\begin{center}
		{\includegraphics[width=1.7\columnwidth,keepaspectratio]
			{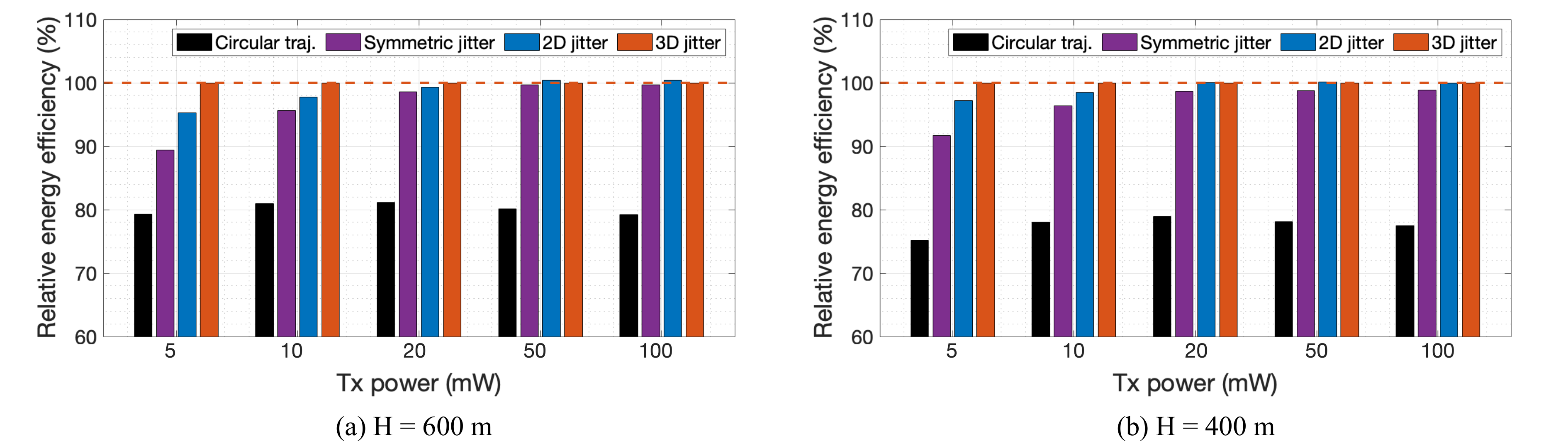}%
			\caption{Relative energy efficiency of trajectories under varying transmit power and altitude conditions.}
			\label{fig12}
		}
	\end{center}
	\vspace{-12pt}
\end{figure*}

\subsection{Comparison with Conventional Jitter Models}

To illustrate the significance of integrating generalized jitter characteristics into trajectory design, we evaluate the performance of our trajectory optimization algorithm using various models of pointing error. In this simulation, the actual UAV jitter is characterized by $[\sigma_\alpha, \sigma_\beta, \sigma_\gamma]=[0.1,1,0.1]$~mrad. The 3D jitter model proposed in this paper fully utilizes the knowledge of $\sigma_\alpha$, $\sigma_\beta$, and $\sigma_\gamma$. However, conventional models do not account for all three DoFs, thus missing information about the actual jittering characteristics. For the two DoF jitter model, the jitter characteristics are best represented by $\sigma_\alpha = \sigma_\beta = 0.711$~mrad and $\sigma_\gamma = 0.1$~mrad~\cite{2018ICC}. For a DoF of 1, implying that $\bold{\Sigma}$ is a scaled identity matrix, the jitter is best represented by $\sigma_\alpha = \sigma_\beta = \sigma_\gamma = 0.583$~mrad. We implement our algorithm to derive the trajectories optimized to pointing error models with one, two, and three DoFs, and then compare their performances under the actual three DoF jitter model. Additionally, a clockwise circular trajectory centered at $[0,-60,H]$~m is used in our simulations as a baseline method. It is important to note that this study is the first to incorporate a jitter model with three DoFs for trajectory design of fixed-wing aircraft.

Our algorithm is designed to maximize energy efficiency, defined as the achievable rate divided by the energy consumption. In Fig.~\ref{fig10}, we compare the average achievable rate of optimized trajectories under varying transmit power levels and different DoF of jitter characteristics. The results indicate that the average spectral efficiency is primarily dependent on the transmit power. Only a slight performance improvement is observed from 1D to 3D jitter models in terms of spectral efficiency. This suggests that the algorithm prioritizes reducing energy consumption over significantly enhancing spectral efficiency. As illustrated in Fig.~\ref{fig11}, the reduction in power consumption is more pronounced, particularly in regions of low transmit power where the algorithm effectively exploits detailed jitter characteristics to minimize power usage. Although using a circular trajectory without optimization might seem intuitive, it consistently shows the lowest performance on both metrics. Furthermore, at lower altitudes, the algorithm tends to focus more on power saving within the power-rate tradeoff. This emphasis on energy conservation is due to the higher average received power at lower altitudes, which lowers the gradient of the rate with respect to changes in trajectory.

In Fig.~\ref{fig12}, we evaluate different jitter models from the perspective of energy efficiency. Across most power and altitude conditions, incorporating multiple DoFs into the jitter model results in a more effectively optimized trajectory design. We set the energy efficiency of the 3D jitter case as the benchmark at 100$\%$ and present the relative values of energy efficiency under different jitter models. Compared to the conventional 1D symmetric jitter model, incorporating a 3D jitter model yields approximately a 10$\%$ performance improvement, especially noticeable in the low transmit power region.

	\section{Conclusion}

In this paper, we first proposed a generalized pointing error model for fixed-wing UAVs in FSO communications, considering the 3D jitter of UAVs. Based on this, we developed a trajectory optimization algorithm to maximize energy efficiency while meeting various constraints such as speed, acceleration, and elevation angles. To solve the complex non-convex problem, we employed the iterative methods and derive an approximation for the constraint associated with the pointing error. Our results emphasize the importance of incorporating the UAV-specific 3D jittering model for accurate performance measurement and system optimization in NTN scenarios. In our simulation, strategically adjusting the flight trajectory significantly reduces the impact of UAV jitter on communication performance, achieving up to a 11.8$\%$ improvement in energy efficiency. This work provides valuable insights into FSO communications performed by fixed-wing UAVs in aerial networks and establishes a basis for further research investigating its applicability in various communication scenarios.

\appendices
\section{Proof of Theorem~\ref{theo1}}
\label{appen1}
First, we apply small angle approximation ($\theta_p\approx\sin\theta_p$) to~(\ref{pedef}). Then, we obtain
\begin{equation}
\label{thetasquare}
\begin{aligned}
\theta_p^2&=\bold{x}^T\bold{A}\bold{x},
\end{aligned}
\end{equation}
with $\bold{A}$ defined as~(\ref{quadraticaa}). The random vector $\bold{x}$ can be substituted by $\bold{x}=\bold{\Sigma}^{1/2}(\bold{\Sigma}^{-1/2}\bold{x})$, where $\bold{\Sigma}^{-1/2}\bold{x}$ is a random vector with iid unit  Gaussian distributed elements. We conduct eigenvalue decomposition as $\bold{\Sigma}^{1/2}\bold{A}\bold{\Sigma}^{1/2}=\bold{P}^T\bold{\Lambda}\bold{P}$, where $\bold{\Lambda}=\bold{\text{diag}}(\lambda_1,\lambda_2,\lambda_3)$. Since ${rank}(\bold{A})=2$ and $\bold{\Sigma}^{1/2}\bold{A}\bold{\Sigma}^{1/2}$ is positive-semidefinite, we can assume that $\lambda_1\geq\lambda_2\geq\lambda_3=0$.
%A is positive semidefinite because there is a unique A=BB where B is a positive semidefinite
So far, we obtain
\begin{equation}
\label{matrixform}
\begin{aligned}
\theta_p^2&=(\bold{\Sigma}^{-1/2}\bold{x})^T\bold{P}^T{\bold{\Lambda}}\bold{P}(\bold{\Sigma}^{-1/2}\bold{x}).
\end{aligned}
\end{equation}
Then, $\bold{P}\bold{\Sigma}^{-1/2}\bold{x}$ can be simplified into $\bold{P}\bold{\Sigma}^{-1/2}\bold{x}=\bold{z}'$ where $\bold{z}'=\left[\zeta_1,\zeta_2,\zeta_3\right]^T$ is an iid Gaussian random vector. Finally, from
\begin{equation}
\label{finalergodicno}
\begin{aligned}
\theta_p^2=\bold{z}'^T\bold{\Lambda}\bold{z}',
\end{aligned}
\end{equation}
we obtain~(\ref{pervs}) by the definition of the Hoyt distribution~\cite{200902EL}.

\section{Proof of Lemma~\ref{theo2}}
\label{appen2}

The term $\mathbb{E}[\log\Gamma]$ can be divided into the sum of each of the channel terms. Since $\Gamma=\frac{e{P}^2}{2\pi\sigma^2}$, it is expressed as
\begin{equation}
\label{finalergodicno}
\begin{aligned}
\mathbb{E}[\log\Gamma]=&\log\left(\frac{e{h_\ell}^2{R}^2{P_\text{T}}^2}{2\pi\sigma^2}\right)\\
&+2\mathbb{E}[\log{h_a}]+2\log{h_\ell}+2\mathbb{E}[\log{h_p}],
\end{aligned}
\end{equation}
where $h_a$ and $h_p$ are the random variables. We obtain
\begin{equation}
\label{loghaapp}
\begin{aligned}
\mathbb{E}[\log{h_a}]=-2\sigma_I^2,
\end{aligned}
\end{equation}
from~(\ref{fading}) and the property of the log-normal distribution, and
\begin{equation}
\label{loghlapp}
\begin{aligned}
\log{h_\ell}=-\sigma_\text{B}z,
\end{aligned}
\end{equation}
from~(\ref{loss}). From~(\ref{pointingloss}) and (\ref{peexpect}),
\begin{equation}
\label{loghpapp}
\begin{aligned}
\mathbb{E}[\log{h_p}]=&\log\left({\frac{a^2}{2\sigma_\text{div}}}\right)-\log{z}-\frac{\lambda_1+\lambda_2}{2{z^2}\sigma_\text{div}^2},
\end{aligned}
\end{equation}
By substituting (\ref{loghaapp}), (\ref{loghlapp}), (\ref{loghpapp}) into~(\ref{finalergodicno}), we obtain~(\ref{capacitytheo}).

\section{Proof of Theorem~\ref{theo3}}
\label{appen3}

We apply first order Taylor approximation to the both sides of~(\ref{nc2}). The LHS of the~(\ref{ca6}) is achieved by substituting $S_k=S_k^{(p)}+(S_k-S_k^{(p)})$ and $U_k=U_k^{(p)}+(U_k-U_k^{(p)})$. The RHS is a function of $\bold{s}[k]$, $\bold{v}[k]$, and $\bold{a}[k]$ since $\hat{\bold{u}}[k]$ can be represented as
\begin{equation}
\label{hatuk}
\begin{aligned}
\hat{\bold{u}}[k]=
\begin{bmatrix}
s_x\cos\theta+s_y\sin\theta\\
-s_x\cos\phi\sin\theta+s_y\cos\phi\cos\theta+s_z\sin\phi\\
-s_x\sin\phi\sin\theta+s_y\sin\phi\cos\theta+s_z\cos\phi
\end{bmatrix},
\end{aligned}
\end{equation}
and $\theta$, $\phi$ are the function of $\bold{v}[k]$ and $\bold{a}[k]$. Through the chain rule, we obtain the first order Taylor approximation $\hat{\bold{u}}[k]\approx\hat{\bold{u}}^{(p)}[k]+\Delta\hat{\bold{u}}[k]$, which is a function of ${\Delta}s_x,{\Delta}s_y,{\Delta}v_x,{\Delta}v_y,{\Delta}a_x,{\Delta}a_y$, where~(\ref{hatup})-(\ref{cosphi}). Also through the chain rule, the linear approximation below is possible:
\begin{equation}
\label{taylorquadsqrt}
\begin{aligned}
&\sqrt{{\hat{\bold{u}}}^T\bold{D}\hat{\bold{u}}}\approx\left({\hat{\bold{u}}^{(p)}[k]}^T\bold{D}\hat{\bold{u}}^{(p)}[k]\right)^\frac{1}{2}\\
&\quad+\left({\hat{\bold{u}}^{(p)}[k]}^T\bold{D}\hat{\bold{u}}^{(p)}[k]\right)^{-\frac{1}{2}}\hat{\bold{u}}^{(p)}[k]\bold{D}\Delta\hat{\bold{u}}[k].
\end{aligned}
\end{equation}

\section{Proof of Lemma~\ref{theo4}}
\label{appen4}
Let $\mathcal{F}$ be a function defined as follows:
\begin{equation}
\label{function}
\begin{aligned}
\mathcal{F}(\bold{x})=\log(\norm{\bold{x}}+H^2)
\end{aligned}
\end{equation}
The Hessian of $\mathcal{F}$ is
\begin{equation}
\label{hessian}
\begin{aligned}
&\bold{H}_\mathcal{F}(\bold{x})=\frac{1}{2(\bold{x}^T\bold{x}+H^2)^2}\left[(\bold{x}^T\bold{x}+H^2)\bold{I}-2\bold{x}\bold{x}^T\right],
\end{aligned}
\end{equation}
where $\bold{I}$ is the identity matrix. Let us prove that $\pi/4\leq\arctan(H/\sqrt{s_x^2+s_y^2})\leq\pi/2$ is a sufficient condition to the convexity of $\log({\sqrt{s_x^2+s_y^2+H^2}})$. Let $\bold{s}'=[s_x,s_y]^T$. The condition can be equivalently represented as $\bold{s}^T\bold{s}\leq{H^2}$. The eigenvalues of $\bold{H}_\mathcal{F}(\bold{s'})$ are as follows:
\begin{equation}
\label{semidefinitepf}
\begin{aligned}
\lambda_{\mathcal{F},1}=\frac{1}{2({\bold{s'}}^T\bold{s'}+H^2)},\,\lambda_{\mathcal{F},2}=\frac{-\bold{s'}^T\bold{s'}+H^2}{2(\bold{s'}^T\bold{s'}+H^2)^2}.
\end{aligned}
\end{equation}
Under the condition of $\bold{s}^T\bold{s}\leq{H^2}$, $\bold{H}_\mathcal{F}(\bold{s'})$ is a positive semidefinite matrix. Therefore, $\frac{1}{2}\mathcal{F}(\bold{s})$ is convex, which guarantees convexity of the constraint~(\ref{nc3}).

\bibliographystyle{IEEEtran}
		\bibliography{Moon_trajectory}


%\vfill


\begin{IEEEbiography}[{\includegraphics[width=1in,height=1.25in,clip,keepaspectratio]{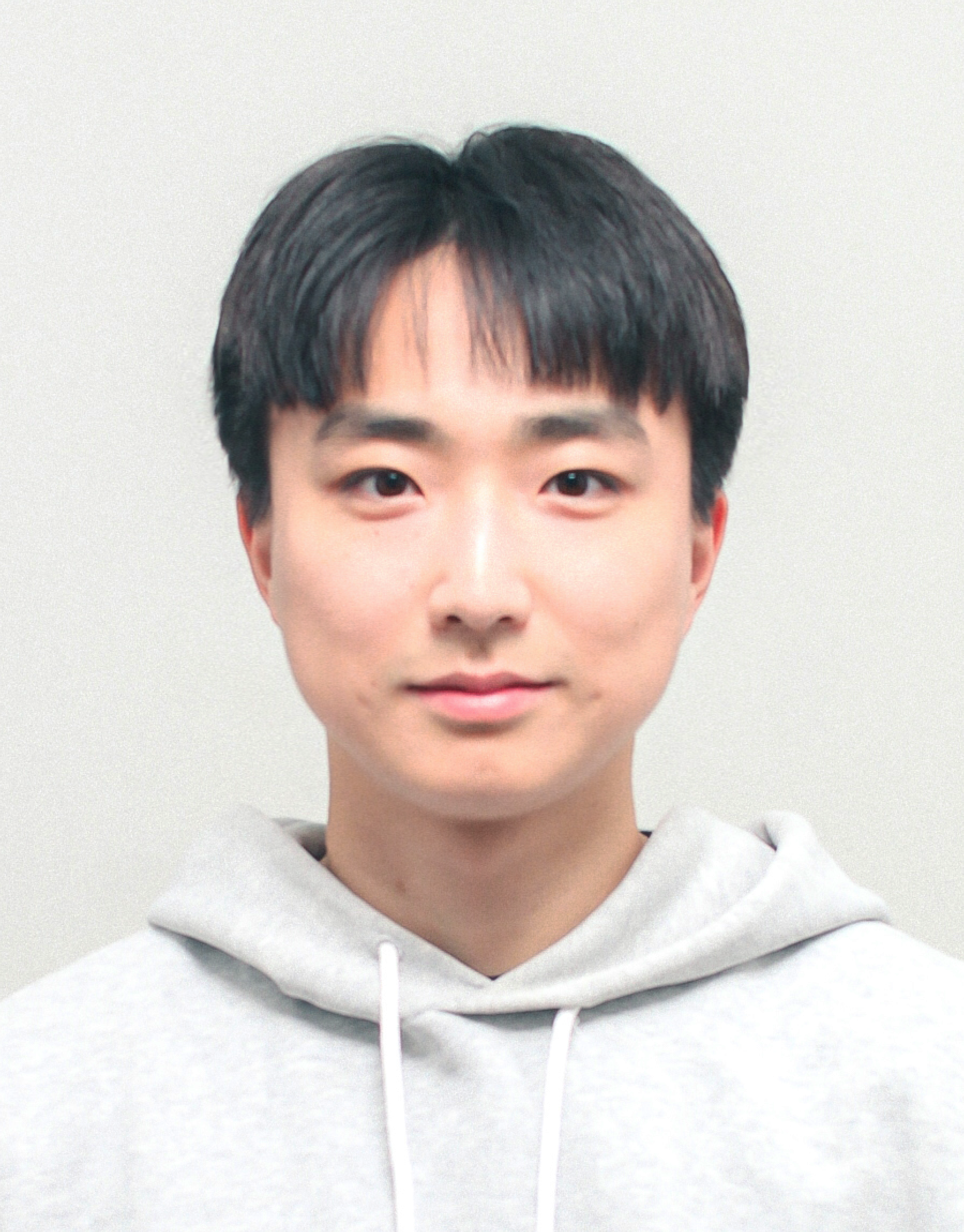}}]{Hyung-Joo Moon} (Student Member, IEEE) received the B.S. degree from the School of Integrated Technology, Yonsei University, South Korea, in 2019, where he is currently pursuing the Ph.D. degree. His research interests include performance analysis and system optimization for emerging technologies in 6G non-terrestrial network (NTN).
\end{IEEEbiography}

\begin{IEEEbiography}[{\includegraphics[width=1in,height=1.25in,clip,keepaspectratio]{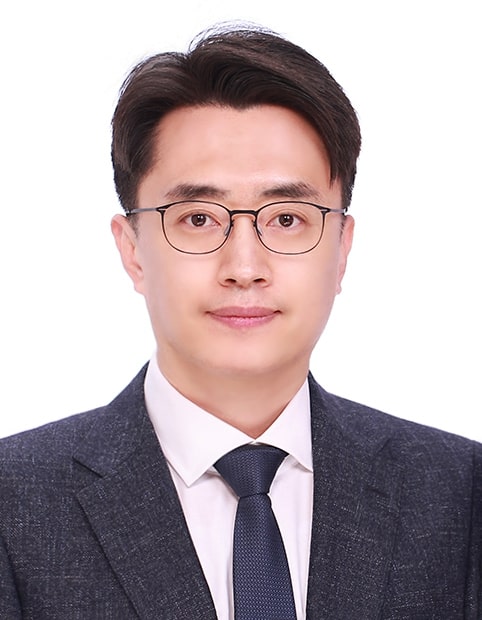}}]{Chan-Byoung Chae} (Fellow, IEEE) received the Ph.D. degree in electrical and computer engineering from The University of Texas at Austin (UT), USA in 2008.

Prior to joining UT, he was a Research Engineer at the Telecommunications Research and Development Center, Samsung Electronics, Suwon, South Korea, from 2001 to 2005. He is currently an Underwood Distinguished Professor with the School of Integrated Technology, Yonsei University, South Korea. Before joining Yonsei University, he was with Bell Labs, Alcatel-Lucent, Murray Hill, NJ, USA, from 2009 to 2011, as a Member of Technical Staff, and Harvard University, Cambridge, MA, USA, from 2008 to 2009, as a Post-Doctoral Research Fellow.

Dr. Chae was a recipient/co-recipient of the Ministry of Education Award in 2024, the KICS Haedong Scholar Award in 2023, the CES Innovation Award in 2023, the IEEE ICC Best Demo Award in 2022, the IEEE WCNC Best Demo Award in 2020, the Best Young Engineer Award from the National Academy of Engineering of Korea (NAEK) in 2019, the IEEE DySPAN Best Demo Award in 2018, the IEEE/KICS Journal of Communications and Networks Best Paper Award in 2018, the IEEE INFOCOM Best Demo Award in 2015, the IEIE/IEEE Joint Award for Young IT Engineer of the Year in 2014, the KICS Haedong Young Scholar Award in 2013, the IEEE Signal Processing Magazine Best Paper Award in 2013, the IEEE ComSoc AP Outstanding Young Researcher Award in 2012, and the IEEE VTS Dan. E. Noble Fellowship Award in 2008.

Dr. Chae has held several editorial positions, including Editor-in-Chief of the IEEE Transactions on Molecular, Biological, and Multi-Scale Communications, Senior Editor of the IEEE Wireless Communications Letters, and Editor for the IEEE Communications Magazine, IEEE Transactions on Wireless Communications, and IEEE Wireless Communications Letters. He was an IEEE ComSoc Distinguished Lecturer from 2020 to 2023. He is an IEEE Fellow and NAEK Fellow.
\end{IEEEbiography}

\begin{IEEEbiography}[{\includegraphics[width=1in,height=1.25in,clip,keepaspectratio]{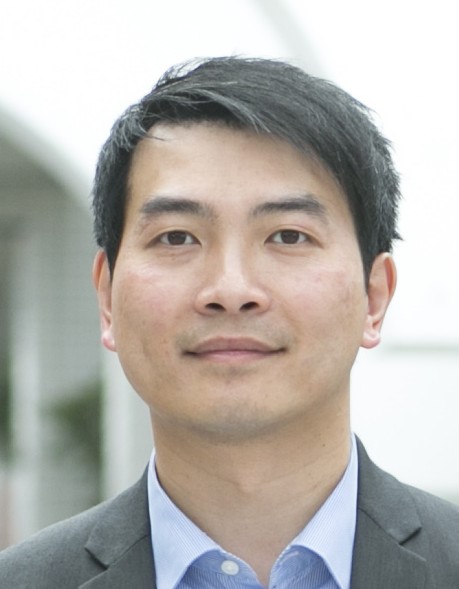}}]{(Kit) Kai-Kit Wong} (Fellow, IEEE) received the BEng, the MPhil, and the PhD degrees, all in electrical and electronic engineering, from the Hong Kong University of Science and Technology, Hong Kong, in 1996, 1998, and 2001, respectively. After graduation, he took up academic and research positions at the University of Hong Kong, Lucent Technologies, Bell-Labs, Holmdel, the Smart Antennas Research Group of Stanford University, and the University of Hull, UK. He is Chair in Wireless Communications at the Department of Electronic and Electrical Engineering, University College London, UK.

His current research centers around 5G and beyond mobile communications. He is a co-recipient of the 2013 IEEE Signal Processing Letters Best Paper Award and the 2000 IEEE VTS Japan Chapter Award at the IEEE Vehicular Technology Conference in Japan in 2000, and a few other international best paper awards. He is Fellow of IEEE and IET and is also on the editorial board of several international journals. He served as the Editor-in-Chief for IEEE Wireless Communications Letters between 2020 and 2023.
\end{IEEEbiography}

\begin{IEEEbiography}[{\includegraphics[width=1in,height=1.25in,clip,keepaspectratio]{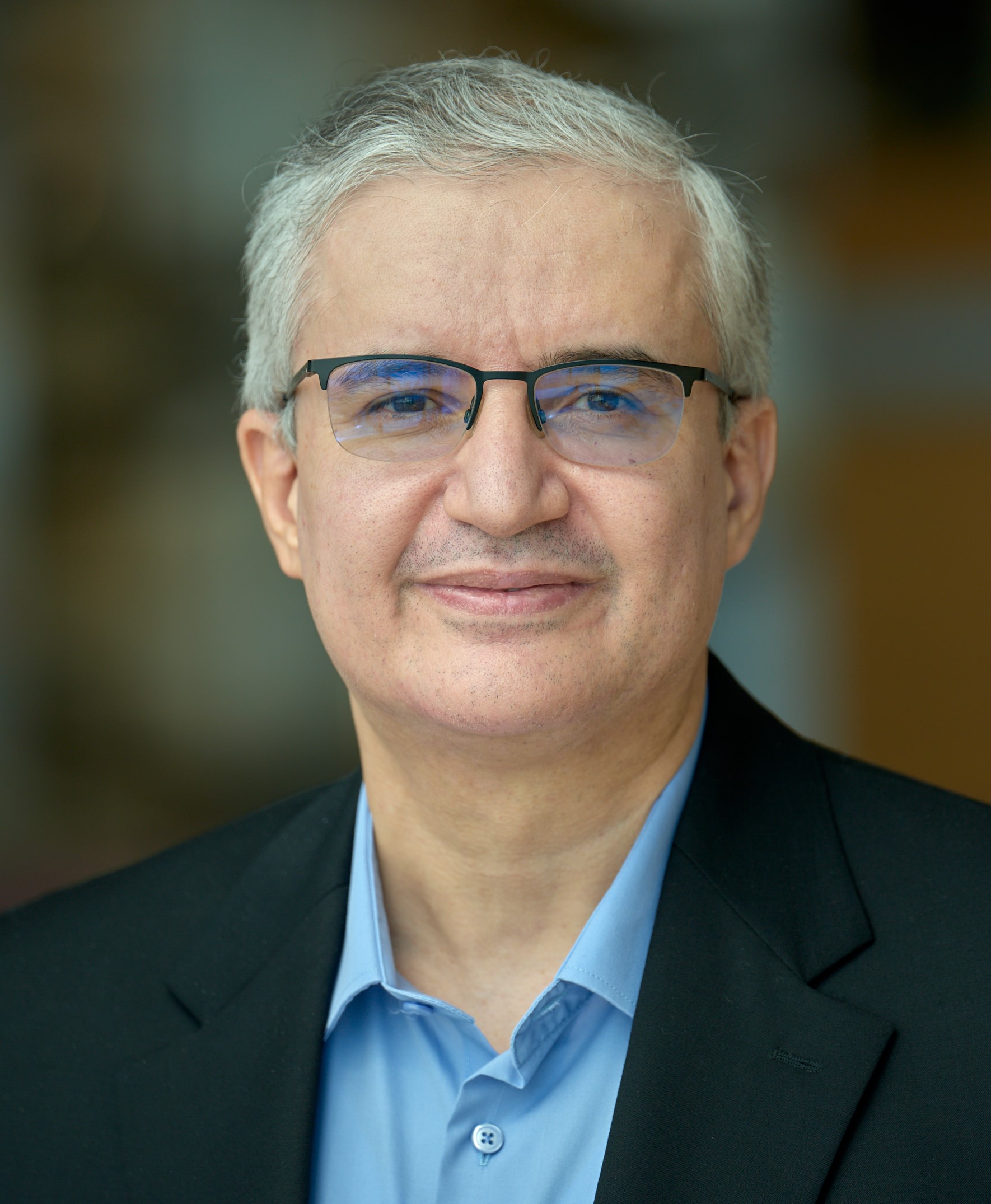}}]{Mohamed-Slim Alouini} (Fellow, IEEE) was born in Tunis, Tunisia. He received the Ph.D. degree in Electrical Engineering from the California Institute of Technology (Caltech), Pasadena, CA, USA, in 1998. He served as a faculty member in the University of Minnesota, Minneapolis, MN, USA, then in the Texas A\&M University at Qatar, Education City, Doha, Qatar before joining King Abdullah University of Science and Technology (KAUST), Thuwal, Makkah Province, Saudi Arabia as a Professor of Electrical Engineering in 2009. His current research interests include the modeling, design, and performance analysis of wireless communication systems.
\end{IEEEbiography}

\end{document}